\documentstyle[aas2pp4]{article}
\input colordvi
\newcommand{\ltwid}{\mathrel{\raise.3ex\hbox{$<$\kern-.75em\lower1ex\hbox{$\sim$
}}}}
\newcommand{\gtwid}{\mathrel{\raise.3ex\hbox{$>$\kern-.75em\lower1ex\hbox{$\sim$
}}}}
\newcommand{\bd}{\begin{description}}
\newcommand{\ed}{\end{description}}
\newcommand{\s}{\scriptscriptstyle}
\begin{document}
\title{Probability of Detecting a Planetary Companion during a
Microlensing Event}
\author{S. J. Peale}
\affil{Dept. of Physics\\
University of California\\
Santa Barbara, CA 93106\\
peale@io.physics.ucsb.edu}
\begin{abstract}
The averaged probability of detecting a planetary companion of a lensing
star during a microlensing event toward the Galactic center when the
planet-star mass ratio $q=0.001$ is shown to have a maximum exceeding
10\% at an orbit semimajor axis near 1.5 AU for a uniform distribution
of impact parameters. This peak value is 
somewhat lower than the maximum of 17\% obtained by Gould and Loeb
(1992), but it is raised to more than 20\% for a distribution of
source-lens impact parameters that is determined by the efficiency of
event detection.  Although these probabilities, based on a signal to
noise (S/N) detection criterion, are model and assumption dependent, the
fact that they change in predictable ways as functions of the orbit
semimajor axes but remain robust for plausible variations of all the
relevant galactic parameters implies that they are representative of
real values. In addition, the averaging procedures are carefully
defined, and they determinine the dependence of the detection
probabilities on several properties of the Galaxy.  The probabilities
for other planet-star mass ratios can be estimated from an approximate
scaling of $\sqrt{q}$.  A planet is assumed
detectable if the perturbation of the single lens light curve exceeds
$2/(S/N)$ sometime during the event, where it is understood that at
least 20 consecutive photometric points during the perturbation are
necessary to confirm the detection.  $S/N$ is the instantaneous value
for the amplified source.  Two meter 
telescopes with 60 second integrations in I-band with high time resolution
photometry throughout the duration of an ongoing event are assumed.
The probabilities are derived as a function of semimajor axis $a$ of
the planetary orbit, where the peak probability occurs where $a$ 
is approximately the mean Einstein ring radius of the distribution of
lenses along the line of sight.  The probabilities remain significant
for $0.6\ltwid a\ltwid 10$ AU.  Dependence of the detection
probabilities on the lens mass function, luminosity function of the
source stars as modified by extinction, distribution of source-lens
impact parameters, and the line of sight to the 
source are also determined, and the probabilities are averaged over
the distribution of the projected position of the planet onto the lens
plane, over the lens mass function, over the distribution of impact
parameters, over the distribution of lens and sources along the line
of sight and over the $I$-band luminosity function of the sources
adjusted for the source distance and extinction.  The probability for
a particular impact parameter and particular source $I$ magnitude but
averaged over remaining degenerate parameters also follows from the
analysis.  In this latter case, the extraction of the probabilility as
a function of $a$ for a particular $q$ from the empirical data from a 
particular event is indicated.
\end{abstract}
\section{Introduction \label{sc:intro}}
There are several searches for extrasolar planets using radial
velocity techniques that together have discovered more than 50 planets
with minimum masses less than 13 Jupiter masses ($M_{\s J}$) ({\it
e.g.}, Marcy, Cochran \& Mayor 2000).  Astrometric searches so far
have been unsuccessful, but new astrometric satellites (Full-sky
Astrometric Mapping Explorer (FAME),  Space Interferometry Mission
(SIM), Global Astrometric Interferometer for Astrophysics (GAIA))
with precisions approaching micro-arcseconds (Lattanzi {\it et al.}
2000) and the Keck Interferometer (Swain {\it et al.} 2000), which may
do almost as well, should soon add additional planets.  Transit
searches have yet to discover a new planet, but a planet discovered
by radial velocity techniques has been successfully observed in
transit (Charbonneau {\it et al.} 2000; Henry {\it et al.}
2000). Continuing transit searches should lead to more discoveries.
All of these search techniques except astrometry favor the discovery
of close, massive planets, and all are restricted to planetary periods
that are only slightly greater than the time the program has been in 
operation or less.  The one astrometric program with a long time base
(Gatewood, 1991) but no detected planets  has too few target stars to be
definitive.  The program initiated by G. Marcy, while having a long
time base, has reached the necessary precision of 3 m/sec to detect distant
Jupiters and Saturns only a few years ago, but the current
program has a sufficient number of stars to constrain the statistics
of distant Jupiters similar to the one in our own solar system in
about 5 years (G. Marcy, D. Fischer, private communications, 2000).

Mao and Paczy\'nski (1991) pointed out that a planetary companion
of a star acting as a gravitational lens amplifying the light of a more
distant star (microlensing)  could perturb the otherwise smooth light
curve in an easily observable way and thereby reveal its
presence. There are many examples in the literature of microlensing 
light curves perturbed by a planet along with descriptions of the
single lens amplification ({\it e.g}. Peale, 1997). The information
about the planet from the perturbed light curve alone is usually
limited to the planet-star mass ratio $q$ and the projected
separation of the planet from the star in terms of the Einstein ring
radius, 
\begin{eqnarray}
R_{\s E}&=&\sqrt{\frac{4GM}{c^2}\frac{D_{\s OL}(D_{\s OS}-D_{\s OL})}
{D_{\s OS}}}, \nonumber\\
&=&\sqrt{\frac{4GMD_8}{c^2}\frac{z(\zeta-z)}{\zeta}},\label{eq:re}
\end{eqnarray}
where $G$ is the gravitational constant, $M$ is the mass of the
lensing star, $c$ is the velocity of light, $D_{\s OL}$ and $D_{\s
OS}$ are the observer-lens and observer-source distances respectively,
$z=D_{\s OL}/D_8$, and $\zeta=D_{\s OS}/D_8$ with $D_8=8$ kpc being
the distance to the center of the Galaxy. Statistically, the projected
star-planet separation is a measure of the semimajor axis of the
planetary orbit (Peale, 1997).   

Part of the search for microlensing events has been
directed toward the Galactic bulge, where the high concentration of
stars in the telescope field allowed a reasonable rate of event
detection and where more than 300 events have, in fact, been detected
(http://darkstar.astro.washington.edu;
http://bulge. princeton.edu/\~ogle/;
http://www.lal.in2p3.fr/recherche /eros/erosa.html). 
Gould and Loeb (1992) showed that the probability of detecting a
planet with $q=0.001$ during a microlensing event approached 17\%
if the planet's projected separation from the star was comparable to
the Einstein ring radius $R_{\s E}$.   A planet was
assumed detectable if the perturbation exceeded 5\% of the
instantaneous flux of the unperturbed single lens light curve. The
probability was averaged over a uniform distribution of lenses along
the line of sight (LOS) to a source at the Galactic center.  The averaged
position of the lens was thus halfway to a source at the galactic
center, where $R_{\s E}\approx 4.1$ AU for a solar mass star meant
that a Jupiter would have nearly the maximum probability of detection. 

This high probability of detecting a planetary companion
during a microlensing event generated an enthusiastic advocacy for
establishing an intensive, dedicated microlensing search program
from the ground (Beichman {\it et al.}, 1996). Initially more modest
microlensing search programs (PLANET (Probing Lensing Anomalies
NETwork), Albrow {\it et al.} 1998, and MPS (Microlensing Planet
Search), Rhie {\it et al.} 2000) have progressed to different
degrees of maturity and are beginning to place meaningful constraints
on the occurrence of Jupiter mass planets around M dwarfs (Albrow {\it
et al.} 2000).  The enthusiasm was reinforced by the special status
noted for very high magnification events, where the probability of
detecting a planet is unity for a range of projected planet-star
separations in the lensing zone surrounding the Einstein ring radius
(Griest and Safizadeh 1998).  The advantage of a microlensing search
over the other search techniques is that the planets in long period
orbits could be discovered  without watching for a decade, and
statistics would start accumulating immediately.  The disadvantage, of
course, is that the lenses are so distant that there could be no
followup studies of the planet after the event. But the knowledge of
the frequency of occurrence of planetary systems and the need to know
these statistics rapidly for planning future planetary search missions
are sufficient to keep microlensing as a most important search
technique. 

This work develops a probability of planet detection during
microlensing events that is based on the photometric signal-to-noise
ratio $(S/N)$ for the amplified source, with the goal of determining
the dependence of this probability on plausible variations in Galactic
parameters while establishing robust averaging techiques over the
distributions of Galactic parameters. The probability is determined as
a function of the planet 
orbit semimajor axis for a planet-star mass ratio of $q=0.001$, and an
approximate scaling law for other mass ratios is indicated. From some
points of view, it might be more desirable to express the
probabilities for planetary masses instead of the planet-star mass
ratio. However, the microlensing calculations are naturally expressed
in terms of $q$, and any attempt to extract a probability in terms of
the planetary masses would have increased already lengthy calculations
by an order of magnitude. As it is debatable whether mass or mass
ratio is the more interpretable quantity, the small return is not
worth the considerable extra effort.  To determine a dependence of the
detection probabilities on Galactic parameters, we must necessarily
average the probabilities over distributions of quantities, such as
lens mass, that remain inherently unknown and over other quantities
such as source-lens impact parameter that are not {\it a priori}
known.  Details of the dependence of the detection probability on most
of the parameters involved are shown as these averages are calculated.
 In this process we  
show that the $S/N$ criterion leads to maximum averaged probabilities
of detection that are comparable to or even greater than that obtained
by Gould and Loeb for plausible assumptions of Galactic parameters,
event geometries and observing telescopes.  Perhaps more importantly,
these probabilities remain robust as the parameters are varied within
reasonable ranges.  We should emphasize to the reader that the  
probability is not that of detecting a planet during any random event,
but only the probability of detection if the lens has a planetary
companion. Hence, the success of any search program depends on the
frequency of planetary occurrence, but null results will be
meaningful in the sense that if the planets are there, they will be
detected with reasonable probability.   

An $S/N$ criterion has been used previously to determine empirical
planet detection probabilities for an individual event  (Gaudi and
Sackett, 2000; Albrow, {\it et al.} 2000a; Albrow {\it et al.} 2000b).
A best fit single lens light curve for a point source-point lens model
is determined from a given event data set by minimizing $\chi_{\s
PSPL}^2$ with free parameters $t_0,\,t_{\s E},u_{\s min},\,F_0,\,f_{\s
B}$. The time of minimum separation between source and lens is $t_0$;
the event time scale $t_{\s 
E}=R_{\s E}/v_{\perp}$, where $v_{\perp}$ is the relative transverse
velocity of source and lens projected onto the lens plane; the
source-lens impact parameter is $u_{min}=\gamma_{min}/\gamma_{\s E}$,
with $\gamma_{\min}$ being the minimum angular separation of source
and lens during the event and $\gamma_{\s E}=R_{\s E}/D_{\s OL}$; 
the unamplified flux density from the source is $F_0$; and the ratio
of the blended flux density from unresolved stars to the source flux density
is $f_{\s B}$. To check if a binary lens gives a better fit to the data,
one assumes a fixed planet-lens mass ratio $q$, a fixed projected planet
separation $b=r/R_{\s E}$, and a projected source trajectory through
the lens plane inclined at angle $\alpha$ relative to the lens-planet
line and again minimizes $\chi^2$ with the same set of free parameters.
A planet is assumed detected if $\Delta\chi^2(q,x_p,\alpha)=\chi_{\s
PSPL}^2-\chi^2(q,x_p,\alpha)>\Delta\chi^2_{thresh}=100$, where the
latter value is considered sufficiently large to be a secure better
fit to the data (Gaudi and Sackett, 2000).  To determine the
efficiency of detecting a $q,b$ planet for a particular event, the
above process is repeated for a uniform distribution of $0\leq
\alpha\leq  2\pi$, and light curves which would have
$\Delta\chi^2=\chi^2(q,x_p,\alpha)-\chi_{\s PSPL}^2$ greater than some
threshold would indicate that the planet would have been detected.
The detection efficiency (probability)  
\begin{equation}
\epsilon(x_p,q)=\frac{1}{2\pi}\int_0^{2\pi}\Theta[\Delta\chi^2(x_p,q,\alpha)-
\Delta\chi^2_{thresh}]d\alpha \label{eq:epsbq}
\end{equation}
is the fraction of trajectories for which $\chi^2(x_p,q,\alpha)$ exceeds
the threshold value for detection for this $(x_p,q)$, where $\Theta(x)$
is a step function. 
The probability in terms of the orbit semimajor axis in AU and planet
mass $m$ is estimated by assuming representative values of the lens
mass and the lens and source distances. 

This procedure confirms that planet detection 
during a microlensing event has a significant probability for individual
events. We show in Section 7 how $\epsilon(x_p,q)$
can be averaged over the degenerate parameters in the Einstein ring
radius weighted by the distribution of lenses and sources along the
LOS and over the mass function (MF) yielding a probability of
detection $P^{\s\prime}(a,q)$ applicable to that particular event with its
determined source magnitude and normalized impact parameter.  The
Gaudi-Sackett procedure cannot be used to obtain the overall averages
sought here for arbitrary events.  Here we have no data
sets on which to establish a $\chi^2$ criterion, so the reference light
curve will be that of the single lens without the planet, and the
perturbations by the presence of a planet will be relative to
this single lens curve.  Since in the real world one has only best
fit single lens light curves and best fit binary lens light curves for
comparison, both of which tend to suppress any perturbations, our
probabilities will be somewhat over estimated.  This overestimate will
be at least partially compensated by our procedure which tends to
underestimate the probabilities.

We justify below the assumption that a planet  will be detectable
if the fractional perturbation of the light curve and hence the mean of 
a set of points during the perturbation exceeds $2/(S/N)$ for the
magnified source. Implicit in this assumption is that the perturbation
be observed sufficiently long that enough photometric points can be obtained
to establish the path of the photometric deviation with more than 99\%
confidence.  In this sense, the $S/N$ for the entire set of anomalous
points will be $\sqrt{\cal N}$ greater than that for an individual point,
where ${\cal N}$ is the number of photometric points during the perturbation.
We argue below that ${\cal N}=20$ will establish a detection with greater
than 99\% confidence if the true mean of the set of perturbed points
is displaced at least $2/(S/N)$ from the unperturbed single lens light
curve. Our neglect of the non-zero angular size of the source will limit
$q\geq 10^{-4}$, where perturbation durations $(2\sqrt{q}t_{\s E})$
will be a significant fraction of a day.  So for almost all binary geometries
where the planetary perturbation will exceed $2/(S/N)$, the duration
of the perturbation will be sufficient to establish a reliable mean of 
the light curve during the perturbation---especially since the
sampling frequency will most likely be drastically increased during the
perturbation. 

The $S/N$ used as the detection
criterion is developed in Section \ref{sc:sn}, where the $S/N$ from
photon statistics and sky noise is reduced by a constant factor to
account for systematic and other unmodeled noise and thereby approximate two
independent empirical determinations of $S/N$ for real systems. As we
shall need a luminosity function (LF) to determine
the fraction of the sources that are visible at any distance $D_{\s
OS}$, we also introduce in this section the LF in $I$-band  along with
its adjustment for the source distance and extinction thereto. We have
chosen the Holtzman {\it et al.} mass function (MF) for the lenses but
the Zocalli {\it et al.} (2000) LF for the sources. Although Holtzman
{\it et al.} (1998) determine  
an $I$-band LF for the Galactic bulge on the way to deriving their MF,
it spans only about 9 magnitudes. But Zocalli {\it et al.} (2000)
have obtained a deeper bulge LF in $J$-band for the lower main sequence and
have supplemented their main sequence LF with the bulge giant LF
obtained from Tiede {\it et al.} (1995) and Frogel and Whitford (1987)
to obtain a more complete LF for the bulge that spans 15
magnitudes.  The straight line segment representation of the main
sequence part of the Holtzman {\it et al.} $I$-band LF has the same 
slope as the LF given in Section \ref{sc:sn} within the uncertainties,
so we keep the MF of Holtzman {\it et al.} (1998) that was derived
from their LF, since programs incorporating it had been written
earlier and since the MFs of several other studies also have an index
of 1 in the M star region ({\it e.g.} Basu and Rana, 1992; Gould,
Bahcall and Flynn, 1997 if unresolved binaries are accounted for 
(Gould, private communication, 1998)), but use the more complete
$I$-band LF developed from the 
$J$-band LF of Zocalli {\it et al.} (2000) in Appendix A.  Zocalli
{\it et al.} (2000) derive a slightly steeper MF in the M star region
than Holtzman {\it et al.} (1998) (index=1.3 {\it vs} 1). The change of the
index from 1 to 1.3 in the M star region has only a small effect on
the detection probability. We choose to illustrate the effect of
increasing the MF index in the M star region by the relatively large
index of 2.2.   

We introduce the fundamental idea of the probability of detection
based on an $S/N$ criterion in Section \ref{sc:detprob}.  Here contours
of the projected position of the planet onto the lens plane, within
which the fractional perturbation of the single lens light curve
exceeds $2/(S/N)$ in magnitude, are shown to sweep out an area in the
lens plane as the source passes by the lens at an impact parameter
$u_{min}=r_{min}/R_{\s E}$.  The extent of the the area grows with
increasing q, with increasing $S/N$, and with decreasing $u_{min}$. We
shall assume $q=0.001$  for most of the calculations, but we also show
a completely averaged probability as a function of the planet
semimajor axis for $q=0.0001$.  (The $S/N$ determined from Section
\ref{sc:sn} is a function of the $I$-band magnitude of the amplified
source for a 2 meter telescope and 60 second integration times.)  If 
the projected position of the planet is 
anywhere inside this area, the planet will cause a fractional
perturbation of the single lens light curve exceeding $2/(S/N)$
for a time of order $2\sqrt{q}t_{\s E}$ sometime during the event and be
detected ({\it e.g.} Gould and Loeb, 1992). The probability of
detection is just the probability that the projected position of the
planet is within the area swept out by the contours.  (The amplification
of the source for various configurations of the lens-planet binary
system and the generation of contours of constant 
perturbation of the single lens light curve used extensively in this
section are developed in Appendix B.)  Also illustrated in this section
is the reason for the unit probability of detection in the annular
``lensing zone'' region straddling $R_{\s E}$ for high
amplification events (Griest and Safizadeh, 1998).)  

The probability of the projected planet position is developed in
Section \ref{sc:lensplanedist} and Appendix C.  To average over
unknown orbital inclination and planet orbital 
longitude, a planet is assumed to be uniformly distributed over a
sphere of radius $a$, and the probability $F(r)$ that the projected
separation of the planet from the lensing star is between $r$ and
$a$ is determined therefrom (Gould and Loeb, 1992).  $F(r)$ is
expressed in terms of $x_p=r/R_{\s E}$ and $a/R_{\s E}$, as this is
the normalization under which the microlensing amplification is
expressed.  Then $F(r)\rightarrow F(x_p,a,\zeta,z,M)$. The explicit
display of $M,\,z$ and $\zeta$ in 
$F(x_p,a,\zeta,z,M)$ follows from $R_{\s E}$ (Eq. (\ref{eq:re})), and 
the degeneracy of these variables in any microlensing event
makes necessary an average over $z$ from observer to source
weighted by the distribution of lenses along the LOS and an
average over $M$ weighted by the MF to yield
$F(x_p,a,\zeta)$ (Appendix C). The average over $\zeta$
weighted by the distribution of sources along the LOS must
follow an average of the probability over the LF.  $F(x_p,a,\zeta)$
is the probability of finding 
the projected position of the planet between $x_p=r/R_{\s E}$ and
$a/R_{\s E}$ averaged over the lens MF and the distribution of lenses
along the LOS from the observer to the source, and
$-dF(x_p,a,\zeta)/dx_p$ is the probability density. The Zhao  (1996)
bulge model is combined with  the Bahcall-Soneira (1980) disk model to
determine the spatial distribution of lenses.  We have already pointed
out our choice of the Holtzman {\it et al} (1998) MF above.
The probability of detection is now
initially developed for fixed $a$ and $\zeta$ for a particular
source $I$-band apparent magnitude.  Eleven values $0.6\leq a\leq
10.0$ are carried through the remaining steps to obtain a final
averaged probability as a function of $a$.

In Section \ref{sc:procedure}, for $q=0.001$ and given I-band magnitude
of the source, $a$ and 
$u_{min}$, we determine the probability $P(a,\zeta,I,u_{min})$
of detecting a planet of semimajor axis $a$, when the normalized
source distance is $\zeta$, the source magnitude is $I$ and the
source-lens impact parameter is $u_{min}$.  This is accomplished by
determining the boundaries of the area swept out in the lens plane by
the contours of the $2/(S/N)$ 
fractional perturbation of the single-lens light curve  and integrating
the probability density of the planet projection over the area, where
the latter simply uses differences in $F(x_p,a,\zeta)$ for two
pairs of values of $x_p$ defining the boundaries of the area at
stepped intervals along the source trajectory.  $P(a,\zeta
,I,u_{min})$ is determined for a distribution of impact parameters
$0<u_{min}<1$, and the results averaged over $u_{min}$ to yield
$P(a,\zeta,I)$, which is the probability of detection for the given
$q=0.001$, semimajor axis, source distance and source $I$-band
magnitude averaged over the distribution of planetary orbital
inclinations and phase, over the distribution of lenses along the LOS,
over the MF, and over the distribution of $u_{min}$.  These
probabilities are determined for a grid of values of $\zeta,\,I$.
It is also shown here how the peak of the detection probability moves
toward smaller values of $a$  as the source moves closer to the
observer, which reflects its following the mean value of $R_{\s E}$.

For each $\zeta$ the span of likely values of $I$ is determined
according to the LF as modified by the distance to and  extinction of
the source with $I<21$ assumed to be the largest magnitude for which
useful high precision, high time resolution photometry could be
carried out.  [For data sets for which image differencing ({\it e.g.},
Alcock {\it et al.} 2000) or other techniques have not been employed
to pick up many of the events with fainter sources, the number of
usable faint sources will be reduced from that implied by the real LF
with $I<21$. Reducing the fraction of sources with low luminosities
would actually increase the averaged probabilities, since the
brighter stars will have a higher S/N. (See Fig. \ref{fig:aveoverlf}.)]
The $P(a,\zeta,I)$ is then averaged over $I_{min}<I<21$
weighted by the LF for each $\zeta$ to obtain $P(a,\zeta)$ for
each $\zeta$. $I_{min}$ corresponds to the brightest star kept in the
LF; it decreases as the source distance $\zeta$ is decreased or the
extinction is reduced. Illustrations of the dependence of the detection 
probability on source distance and source magnitude for averages over
the LF at each source distance are also given in Section
\ref{sc:procedure}.  The dependence of $I$ on $\zeta$ is why we
delayed averaging over the latter until the average over the LF could
be completed.   

For $0.1<\zeta<1.2$, $P(a,\zeta)$ is averaged over the 
distribution of {\it visible} sources along the line of sight in Section
\ref{sc:paveoverzeta} to yield $P(a)$, where ``visible'' means $I<21$
as discussed above. $P(a)$ is the probability of detecting a
planetary companion of a lens with $q=0.001$ and semimajor axis $a$
(for 11 values of $a$) during a microlensing event averaged over all
of the Galactic and geometric parameters. The
same galactic model used for the distribution of lenses is used for
the distribution  of sources. The lower limit on $\zeta>0$ is chosen
to avoid a singularity in the expressions, where a source at
$\zeta<0.1$ is extremely unlikely in any case. The upper limit is
chosen 1.2 since extinction and blending will make the number of sources
that are usably visible negligibly small at larger $\zeta$. It is shown
here that for a LOS toward Baade's window, the peak probability of
detection exceeds 10\% for $a$ near the mean value of $R_{\s E}$ of
almost 2 AU, if $u_{\s min}$ is distributed uniformly, but it exceeds
20\% for a distribution determined by the observational efficiency for
the events, which has a higher proportion of small $u_{min}$. 

To clarify the series of averages, we write an equation for $P(a)$
that includes all of the averaging operations except the initial one
over the planetary orbital inclination and phase, where the latter is
accomplished simply by the assumption of uniform distribution over a
sphere of radius $a$. 
\begin{displaymath}
P(a)=\int_{0.1}^{1.2}n_{\s S}^{\s\prime}(\zeta)d\zeta
\int_{I_{min}}^{21}LF(I,\zeta,A)dI
\end{displaymath}
\begin{displaymath}
\times\frac{1}{k}\sum_{i=1}^k
\int\!\!\int_{S(I,u_{min}^i)}dS\left(-\frac{d}{dx_p}\right)
\end{displaymath}
\begin{displaymath}
\times\int_0^\zeta\int_{M_{min}}^{M_{max}}n_{\s L}^{\s\prime}(z,M)
\Theta[f(x_p,a,\zeta,z,M)]
\end{displaymath}
\begin{equation}
\times F(x_p,a,\zeta,z,M)dzdM, \label{eq:grandint}
\end{equation}
where $\Theta[f(x_p,a,\zeta,z,M)]$ is a step function whose argument
$>0$ [given in Eq. (\ref{eq:fxpave})] is a condition for the reality of
$F(x_p,a,$ $\zeta,z,M)$; $n_{\s 
L}^{\s\prime}(z,M)$ is the fraction of lenses in volume
$z^2dz\Delta\Omega$ and in mass range $dM$ of those along the LOS to
the source with $\Delta\Omega$ being a representative solid angle in
the field; $S(u_{min}^i,I)$ is the area in the lens plane
corresponding to the particular $u_{min}^i$ and source $I$-band
magnitude within which a planet would cause a
perturbation of the light curve exceeding $2/(S/N)$.  The double
integral on the far right is $F(x_p,a,\zeta)$ so
$-dF(x_p,a,\zeta)/dx_p$ is the probability density, which is
integrated over the area in the next integral to the left to yield
$P(a ,\zeta,I,u_{min}^i)$.  The sum of $P(a,\zeta,I,u_{min}^i)$ over the
$k$ values of $u_{min}$ divided by $k$ gives the average over the impact
parameters to yield $P(a,\zeta,I)$. $LF(I,\zeta,A)$ is the
fraction of sources per unit $I$ magnitude adjusted for distance and
extinction $A$.  This integral over the LF yields
$P(a,\zeta)$. Finally, $n_{\s S}^{\s\prime}(\zeta)$ is the fraction of
{\it visible} sources of those along the line of sight, and the
integral over $\zeta$ yields $P(a)$.

Section \ref{sc:paramdep} shows how the detection probability changes
as various parameters are changed. It decreases as the extinction is
increased or if the detection criterion is increased from $2/(S/N)$ to
$3/(S/N)$.  Changing the LOS changes the distribution of the
probability over the values of $a$ by changing the mean
$R_{\s E}$. Here we show how increasing the fraction of M-type stars
in the MF drives the peak in the probability to smaller values of $a$,
while decreasing the probability for large $a$ and increasing the
probability for small $a$. We have already pointed out above the
conversion of the model independent $\epsilon(x_p,q)$ to the model
dependent $P^{\s\prime}(a,q)$ at the end of Section 7.

The properties of the averaged planetary detection probability are
summarized in Section \ref{sc:summary} where it is also shown how the
detection probability, averaged over everything except the LF, varies
with the source magnitude. Also shown here is the detection
probability as a function of the planetary semimajor axis for a
particular event, where the source magnitude and the impact parameter
are assumed known.   A discussion follows in Section
\ref{sc:discussion}, where the robustness of relatively large
detection probabilities over a wide range of planetary semimajor axes
to variations in the various parameters is emphasized, while the
dependencies on assumptions and other caveats and neglected processes
are pointed out.  The unique aspects of a microlensing search for
planets and the robustness of the probabilities of detection for all
plausible variations in Galactic parameters are additional 
justifications for the continued pursuit of ground based microlensing
searches for planets in spite of the exponentially growing success of
radial velocity searches and proposed space based microlensing
searches. 
\section{Signal to noise ratio. \label{sc:sn}}
The part of $S/N$ appropriate to photometry
of a star that depends on just photon statistics and sky noise can be
written (up to the factor 1/4 in parentheses) 
\begin{eqnarray}
\frac{S}{N}&=&\frac{\phi_{\s star}\sqrt{A_{\s T}Et}}{\sqrt{A_{s}\Phi_{\s
sky}+\phi_{\s star}}}\left(\times\frac{1}{4}\right)\nonumber\\
&=&\frac{314\times 10^{-0.4I}D\sqrt{t}}{\sqrt{2.23\times
10^{-8}+10^{-0.4I}}}\left(\times\frac{1}{4}\right), \label{eq:sn}
\end{eqnarray}
where $\phi_{\s star}$ (photons/${\rm cm^2 sec}$) is the photon flux
density from the source star, $A_{\s T}=\pi D^2/4$ ($\rm cm^2$) is the
telescope aperture area ($D=$ aperture diameter), $E$ is the overall
efficiency of the system, $\Phi_{\s sky}$ (mag/arcsec$^2\rightarrow
{\rm photons/(cm^2 sec\,arcsec^2)}$) is the surface brightness of the
sky, $A_s=\pi s^2/4$ (arcsec$^2$) is the seeing disk ($s=$ seeing
defined as the FWHM of the point spread function) and
$t$ (sec) is the integration time. From the lines just 
preceding Eq. (\ref{eq:li}), the $I$-band flux density from Vega is
$1.56\times 10^{-6}{\rm erg\,cm^{-2}sec^{-1}}$ leading to
$\phi_{\s star}=6.28\times 10^5 10^{-0.4I}$ ${\rm
photons\,cm^{-2}sec^{-1}}$ for magnitude $I$ with a square profile for
the $I$-band filter $0.13\mu{\rm m}$ wide at $0.8\mu{\rm m}$ ($I_{\s
Vega}=0$). The  clear night sky brightness at Cerro Tololo
Interamerican Observatory in $I$-band ranges from 19.2 
to 19.9 magnitudes {$\rm arcsec^{-2}$} (Walker, 1987) from full to new
moon. We choose an average sky brightness of 19.6 magnitudes {$\rm
arcsec^{-2}$}, $s=1.4$ arcsec and $E=0.2$ to yield the final form
of Eq. (\ref{eq:sn}) without the factor (1/4).

Eq. (\ref{eq:sn}) without the factor 1/4 does not yield values of
$S/N$ close to those empirically determined because of
instrument and systematic noise, reduced flux from sky absorption and
effects of other varying observing conditions that have been omitted.
To account in an approximate way for these omitted quantities, we
reduce the coefficient of the $S/N$ from photon statistics and sky
noise  until $S/N$ matches the empirical data of Albrow
{\it et al.}, (1998).  For $D=91\,{\rm cm}$ and $t=300\,{\rm sec}$,
Albrow {\it et al.}  obtain 1\%, 2\%, 7\% photometry for
$I=15,\,17,\,19$ respectively. For $I=19$, substitution of these
values of $D$ and $t$ into Eq. (\ref{eq:sn}) yields $S/N=57$ instead
of a little over 14 for the 7\% photometry found by Albrow {\it et
al.} The empirical datum is matched if we multiply Eq. (\ref{eq:sn})
by the indicated factor of (1/4).  Table 1 compares the $S/N$
determined by Eq. (\ref{eq:sn}) with the empirical values of Albrow
{\it et al.} for the remaining two magnitudes and for empirical values
for a 76 cm telescope at Lick Observatory
(W. Li and A. Filippenko, Private communication, 2000). The Albrow
{\it et al.} data are nearly matched for all the magnitudes, whereas
the Lick data are reproduced for the brighter magnitudes, but less
well for the magnitudes near the limiting value for the telescope.
Eq. (\ref{eq:sn}) as modified is thus a good approximation at least to
the Albrow {\it et al.} data, and reproduces much of the Lick data.  
These comparisons with empirical data encourage the use of
Eq. (\ref{eq:sn}) for our purposes here.  In determining $S/N$ to be
used  in finding the probability of detecting a planetary companion of
a lens during an event, we shall assume $D=200\,{\rm cm}$ and $t=60$
sec, except we shorten the integration time for bright sources to keep
$S/N<200$.  Systematic effects would probably soon dominate photon
statistics for the brighter sources, so $S/N$ would not be improved by
longer integrations. 
\begin{table}[h]
\centering
\begin{tabular}{|c|c|c|} \hline
I mag&Albrow {\it et al.}&Lick\\ \hline
&D=91 cm&D=76 cm\\
&t=300 sec&t=300 sec\\
&s=1.4 as&s=2.24 as\\ \hline
15&100, {\bf 122}&$>80$, {\bf 100}\\ \hline
17&50, {\bf 46}&35, {\bf 35}\\ \hline
18.5&&6, {\bf 13}\\ \hline
19&14, {\bf 14}&2-3, {\bf 9}\\ \hline
\end{tabular}
\caption{Comparison of empirical determinations of $S/N$ with
those determined by Eq: (\ref{eq:sn}) including the factor 1/4.  The
calculated values of $S/N$ are in boldface type next to the empirical values.}
\end{table}

Subjective impressions of examples of precision photometry in the
literature, ({\it e.g.} Charbonneau {\it et al.} 2000), suggest that a
$2\sigma$ perturbation of a single lens light curve should be easily
discerned with a sufficient number of data points, where
$\sigma=1/(S/N)$ is the rms fractional deviation of  points about the
best fit curve ({\it e.g.}, $\sigma=.01$ is 1\% photometry).  What is
``sufficient''  is also subjective, but we can quantify this in terms
of confidence. We consider a microlensing event of duration 30 days
($2t_{\s E}$) perturbed by a planet with $q=0.0001$, where the expected
duration of the perturbation is $\sim 2\sqrt{q}t_{\s E}=0.3$ days
({\it e.g.}, Peale, 1997).  As an example, we assume a worst case true
perturbation of the single lens light curve remains at the minimum $2/(S/N)$
for the entire 0.3 days, where $S/N$ is that for a single observation
of the amplified brightness of the source, which will not change
significantly during the short time of the perturbation.  This example
perturbation is unlike a true planetary perturbation, but its
simplicity and minimal nature will illustrate the viability of our
detection criterion adequately.  We want to
find an estimate of the number of data points necessary to detect such
a perturbation coupled with a measure of the confidence in the
detection. 

\begin{figure}[ht]
\plotone{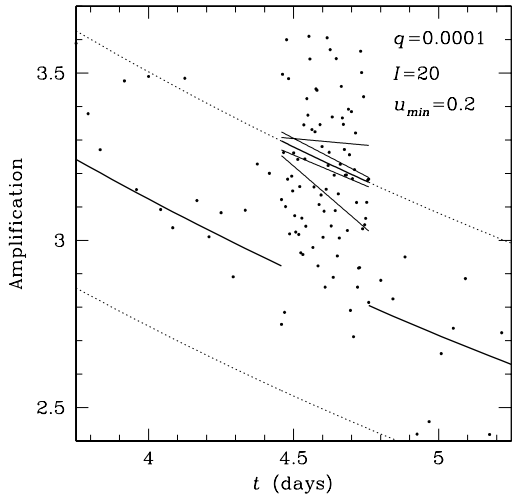}
\caption{Segment of a microlensing light curve for $I=20,\;u_{min}=0.2$,
and $q=0.0001$, where the duration of the planetary perturbation is
$2t_{\s E}\sqrt{q}=7.2$ hours for an event duration of 30 days.
The bold lines represent the true amplification and 
the dotted lines are displaced by a fraction of $\pm 2/(S/N)$ from the
true amplification. The simulated points are gaussian distributed with
$\sigma=1/(S/N)$ The two short thinner lines on
either side of and closest to the bold line during the perturbation are
least squares  
fits to the set simulated data points and to a set where each deviation
is given the opposite sign. The second set of lines further separated
from the true amplification during the perturbation are least square
fits to only 20 data points.  \label{fig:snexample}} 
\end{figure}

In Fig. \ref{fig:snexample}, we
show a segment of a microlensing light curve with $u_{min}=0.2$
for a source magnitude $I=20$ with this minimum assumed perturbation
that occurs about 4.5 days after the time of the peak
amplification.  The solid bold line is the true amplification of the
source, and the dotted curves span the true unperturbed light curve by a
fraction $\pm 2/(S/N)$. $S/N$ as determined by
Eq. (\ref{eq:sn}) has a relatively poor value of about 15 at the
region of the curve where the perturbation occurs.  The simulated data
points are normally distributed with $\sigma$ for the fractional
deviation equal to the local value of $1/(S/N)$.  The sampling
frequency is once per hour outside the perturbation region, but it is
increased to once every 5 minutes during the perturbation with the same
conditions (integration time, telescope, {\it etc}.) determining $S/N$
as for the remaining part of the light curve.  The two short lines
closest to and on either side of the bold line are least squares
fits to a straight line for the 86 simulated data points and to the
same set with opposite signed, equally probable deviations.
The uncertainty $\sigma_{det}$ in the mean position of the normalized
data points 
during the perturbation is $\sigma/\sqrt{\cal N}=1/(\sqrt{86}S/N)$,
which is a measure of the uncertainty in the detection.  The effective
$S/N$ for the detection is more than 9 times that for each individual
point and the detection is essentially certain. That is, with 99\%
confidence, the true amplification is within $3\sigma_{det}=1/(3S/N)$
of the measured mean represented by the midpoint of the either of the
two short lines segments closest to the true light curve during the
perturbation.  The uncertainty is small compared with the magnitude of
the fractional perturbation of $2/(S/N)$.

The other two small line segments separated further from the true
light curve correspond to least squares fits to only 20 points taken
during the perturbation, which is a sampling frequency about 3 times
that for the other parts of the light curve. Here
$\sigma_{det}=1/(\sqrt{20}S/N)$ and there is 99\% confidence that the
true amplification falls within about $2/(3S/N)$ of the midpoint of either
line.  This is still sufficiently less than the $2/(S/N)$ perturbation
that a detection would be secure.  There is a necessary cushion here,
because the entire light curve will only be a best fit model of all of
the points, and the perturbation will be suppressed somewhat during the
fit. If we assume a planetary perturbation of the single lens light
curve that exceeds 
$2/(S/N)$ can be detected with 20 consecutive data points during the
perturbation, confidence in the detection would exceed 99\% even for
the minimum perturbation.  A detection criterion can then be defined
as a fractional perturbation of the single lens light curve that exceeds
$2/(S/N)$ for at least 20 consecutive data points.  Sufficient
duration of a perturbation to satisfy this criterion will obtain for
almost all of the events considered in determining the
probabilities for reasonable sampling rates with the applicable values
of $q$, and duration will not appear explicitly in the calculations
to follow. In this sense, duration is really not part of the detection
criterion used in the calculations, but the constraint of ``20
consecutive data points'' is understood to be satisfied for nearly all
detectable events. Note that for the minimum fractional perturbation of
$2/(S/N)$, some part of the perturbation will probably have passed
before a higher sampling frequency and perhaps longer integration
times were effected to more closely map the perturbed part of the
light curve. Hence, a  detection does not necessarily imply sufficient
definition of the light curve to constrain $q$ adequately.  In
addition, the poorer definition of the slope of the best fit curves
when only 20 data points are available illustrates the necessity of
very frequent sampling if one hopes to characterize the planet with
any certainty.   It is certainly clear from Fig. \ref{fig:snexample}
that an $S/N$ considerably better than 15 is desirable for
characterizing the planet, but the smaller values do not preclude
detecting a $2/(S/N)$ perturbation with 20 data points.  

The $S/N$ used to establish the criterion for
detecting a planetary companion during a microlensing event depends of
course on the apparent brightness of the source star.  The
distribution of the apparent brightness of the source stars at a
particular distance are determined from the following LF.
\begin{eqnarray*}
f_{\s L}(I_8)&=&1.470\times 10^{-9}e^{0.930\,I_{\s 8}}\quad 10.5\le
I_{\s 8}\le 17\\ 
&=&2.575\times 10^{-8}e^{0.762\,I_{\s 8}}\quad 17\le I_{\s 8}\le 18.5
\\ 
&=&1.316\times 10^{-4}e^{0.300\,I_{\s 8}}\quad 18.5\le I_{\s 8}\le
26,\\
&&\qquad\qquad\qquad\qquad\qquad(A5)
\end{eqnarray*}
which represents the fraction of the sources per unit $I$ magnitude
for bulge stars at a distance of 8 kpc.  $I_8$ is the $I$ magnitude at
that distance.  This $I$-band LF is constructed from the $J$-band LF of
Zoccali {\it et al.} (1999) in Appendix A, where the $I$-band is
chosen as the most likely bandpass that will be used in any dedicated
planet search from the ground.  In Eqs. (\ref{eq:lfi}), $10.5<I_8<17$
corresponds to giant stars and $18.5<I_8<26$ to main sequence stars
below the turnoff with the second of Eqs. (\ref{eq:lfi}) bridging the
gap between the two distributions.  The same luminosity function will
be assumed to apply to the disk as well as the bulge, although there
will be fewer giants among the disk stars.  Most of the sources will
be in the bulge in any case, and changes induced in the probability
calculations  by a more refined LF are negligibly small.

For particular LOS, $\zeta$, $q$, $a$ and MF of the
lenses, the probability of detection  is calculated for a few tens of
values of the impact parameter ($0<u_{min}\ltwid 1$)
and the probabilities averaged over the impact parameter
distribution. This is done for approximately 10 different apparent
magnitudes in the range $I_{\rm min}<I<21$.  The maximum value of
$I=21$ is chosen as the dimmest star that can be usefully monitored in
a ground based search for planetary perturbations, whereas $I_{\rm
min}$ is determined by the brightest star in the LF translated to
distance $\zeta$ and increased by the extinction to that distance. The
extinction is handled by assuming that the measured extinction to 
8 kpc along the particular LOS is uniformly distributed from
the observer to 8 kpc.  The transformation of the LF of
Eq. (\ref{eq:lfi}) to one for the apparent magnitude at distance
$\zeta\neq 1$ is effected by substituting  
\begin{equation}
I_8=I_{\rm app}-A_{\s I8}\zeta-5\log{\zeta} \label{eq:i8vsiappar}
\end{equation}
into the LF, where $A_{\s I8}$ is the extinction to 8 kpc. Details are given
in Appendix A. 
\section{Probability of detection \label{sc:detprob}}
\begin{figure}[ht]
\plotone{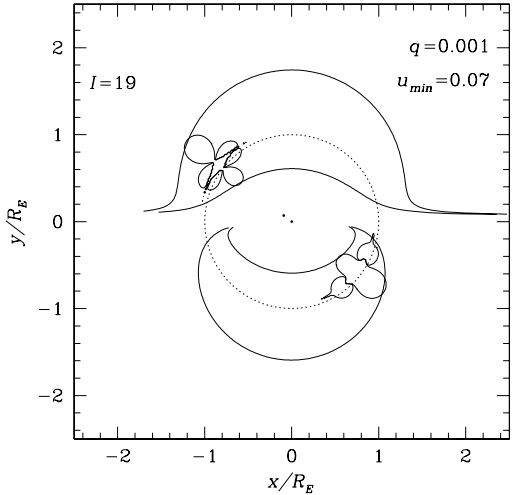}
\caption{Sweep of small contours of $\pm 2/(S/N)$ fractional
perturbation of light 
curve across the lens plane. The large curves are traces of the
extremes of the contours during the event, and they define the area
within which a planet would be detected if it happened to be
there. The source  magnitude is I=19.  The diagram is centered on the
lens with the position of the source corresponding to the contours
displaced a small distance into the second quadrant.  The dotted
circle is the Einstein ring. At the source location shown,
$2/(S/N)=0.0366$, whereas at closest approach $2/(S/N)=0.0283$. The
source moves horizontally across the
diagram. \label{fig:contoursweepiband}}  
\end{figure}

The basic idea behind the detection probability is due to Gould
and Loeb (1992), although we use an $S/N$ criterion for
detection instead of a fixed percentage perturbation of the
microlensing light curve.   Fig. \ref{fig:contoursweepiband} shows loci of
equal amplification for the positions of a planet projected onto the
lens plane  with planet-star mass ratio $q=0.001$. The amplification of
the source is $A=[1\pm 2/(S/N)]A_0$ if the planet is on a contour,
where $A_0$ is
the unperturbed amplification of the source if the planet were not
there. A fractional perturbation of $2/(S/N)$ is assumed  detectable
with $(S/N)$ being  that appropriate to the
instantaneous brightness of the amplified source. Inside the loci
the perturbation will exceed $2/(S/N)$ in magnitude.  The method of
estimating $S/N$ for a given source magnitude is shown in Section
\ref{sc:sn} above.

As the source passes by the lens from left to right in
Fig. \ref{fig:contoursweepiband}, the contours sweep across the lens
plane---waxing and waning as the source gets closer to or further from
the lens and as $S/N$ varies with the instantaneous value of
$A_0$. The large curves trace the extremes of the equal
amplification loci during the traverse of the source across the
Einstein ring, and a planet would be detected at some time during the
event if its projected position were anywhere within either pair of
the large curves.  We adopt the probability of detecting the planet
during the event as the probability that its projected position is
within the area contained by the large curves.  This area expands and
the probability of detection is correspondingly higher as the source
brightness is increased thereby increasing $S/N$ and as the impact
parameter of the source-lens encounter is reduced.  The contours
shrink or expand approximately as $\sqrt{q}$, and the area between
the trace curves and probability of detection are changed
approximately by the same factor. 

\begin{figure}[ht]
\plotone{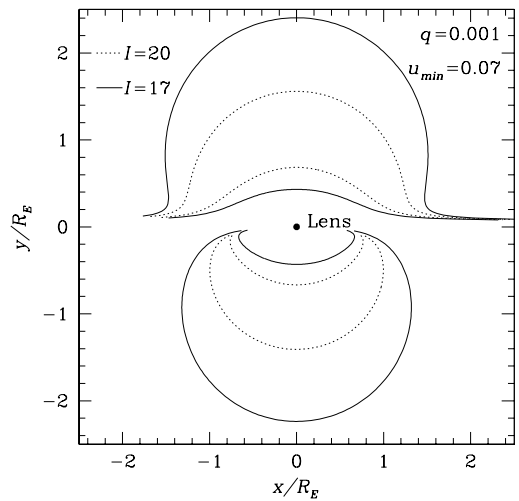}
\caption{Traces of the extremes of the contours of $\pm 2/(S/N)$
fractional perturbation of the light curve for $q=0.001$,
$u_{min}=0.07$ and two stellar I-magnitudes. The area inside the
curves and the probability of detection is higher for brighter
stars.\label{fig:extrcrvsmagcmpreiband}}  
\end{figure}

\begin{figure}[ht]
\plotone{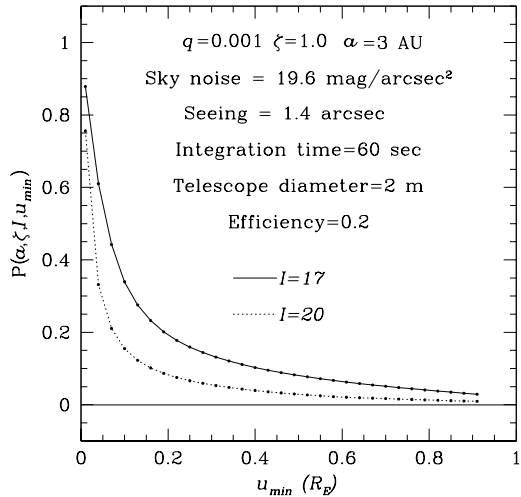}
\caption{Detection probability for $q=0.001,\,\zeta=1.0,\,a=3$ AU
as a function of the impact parameter for two different source
brightnesses.  The probability increases as $u_{min}$ is reduced, and
it is increased for all impact parameters for the brighter
star. \label{fig:probvsubiband}}  
\end{figure}

Fig. \ref{fig:extrcrvsmagcmpreiband} shows how the area is increased with
source brightness. The effect on the probability of this change in
area is shown in Fig. \ref{fig:probvsubiband}, where the detection
probability is increased for brighter stars for all 
$u_{min}$. Fig. \ref{fig:probvsubiband} also illustrates the
increased probability of detection for smaller $u_{min}$.  The
impact parameter of 0.07 in Fig. \ref{fig:contoursweepiband} is
reduced to 0.005 in Fig. \ref{fig:contoursweep2}, which
shows the expanded loci of minimum perturbation for detection and the
expanded curves tracing the extremes of these loci.  One notices first in
Fig. \ref{fig:contoursweep2} that the loci extend beyond the traces of
their extremes along their symmetry axes (the source-lens line), so
that the probabilities 
are slightly underestimated by our adopted scheme.  That is, a
planet would still be detected if it were slightly outside the curves
defining the traces of the extremes of the loci contours.  However,
this effect is most pronounced for the smallest impact parameters
where it makes at most a few percent difference in the calculated
probability, and we average these probabilities over the distribution
of impact parameters, which further reduces the relative contribution
of this extra probability for small impact parameters.  Notice next
that the contours within which the planet would be detected completely
encompass the region around the Einstein ring radius for small impact 
parameters.  This illustrates the result of Griest and Safizadeh
(1998) who found unit probability of planet detection for high
amplification events if the projected position of the planet were
anywhere within a region spanning the Einstein ring radius.

The amplification of a source by a binary star-planet lens is
determined in Appendix B with the analysis in the
complex plane by Witt (1990).  The procedures for determining the
contours of equal amplification and for determining the extremes of
these contours---used in Figs. \ref{fig:contoursweepiband},
\ref{fig:extrcrvsmagcmpreiband} and \ref{fig:contoursweep2}---are also
outlined in this appendix.   
\section{Probability distribution of the planet in the lens plane.
\label{sc:lensplanedist}}
The time average separation of a planet from its primary is $a(1+e^2/2)$,
where $a$ and $e$ are the semimajor axis and eccentricity of the
orbit.  So we can account for the random orbit orientation and random
phase by assuming that the instantaneous position of a planet is
uniformly distributed over the surface of a sphere of this radius.
However, we drop 
the $e^2$ and assume the sphere has radius $a$, where it is understood
that this ``semimajor axis''  is augmented by some assumed mean
orbital eccentricity to represent an averaged separation. 
The cumulative probability that a planet with
this distribution will have a position projected onto the lens plane
between $r\leq a$ and $a$ is (Gould and Loeb, 1992)  
\begin{displaymath}
F(r)=\sqrt{1-\frac{r^2}{a^2}},
\end{displaymath}
or between $x_p=r/R_{\s E}$ and $a/R_{\s E}$,
\begin{eqnarray}
F(x_p,a,\zeta,z,M)&=&\sqrt{1-R_{\s E}^2\frac{x_p^2}{a^2}}\nonumber\\
&=&\sqrt{1-K\frac{M}{a^2}\frac{z(\zeta-
z)}{\zeta}x_p^2},\label{eq:cumprobfxp}
\end{eqnarray}
where $K=4GM_{\s\odot}D_8/(c^2a_{\s\oplus}^2)=65.1279$ with 
$G=$ the gravitational constant, $M_{\s\odot}=$ the solar
mass, $a_{\s\oplus}=1$ AU, $M$ is the lens
mass in units of $M_{\s\odot}$, and $a$ is now measured in AU in the
last line.

We wish to determine a planet detection probability during a
microlensing event in terms of a criterion based on a representative
$S/N$ of the signal.  The latter depends on the apparent
magnitude of the source, which in turn depends on $\zeta$ for the
given luminosity function. We therefore average the cumulative
probability in Eq. (\ref{eq:cumprobfxp}) over the distribution of the
lenses from the observer to a particular source distance $\zeta$ and
simultaneously over the MF of the lenses, since the
lens mass $M$ is embedded in the definition of $R_{\s E}$
(Eq. (\ref{eq:re})). The probability of detection of a planet of given 
$q$ (averaged over the lens density distribution and MF about a star 
lensing a source of given $I$ magnitude at given distance $\zeta$ can
then be averaged over the apparent brightness distribution from the LF
as adjusted for that particular distance.
After the probabilities are averaged over the MF, lens spatial
density distribution and LF, they will then be averaged over the
distribution of {\it visible} sources along the particular line of
sight---all this as a function of the semimajor axis of the planetary
orbit in AU. But first we must average $F(x_p,a,\zeta,z,M)$ over the lens
distribution between observer and source and over the MF. 
\begin{figure}[ht]
\plotone{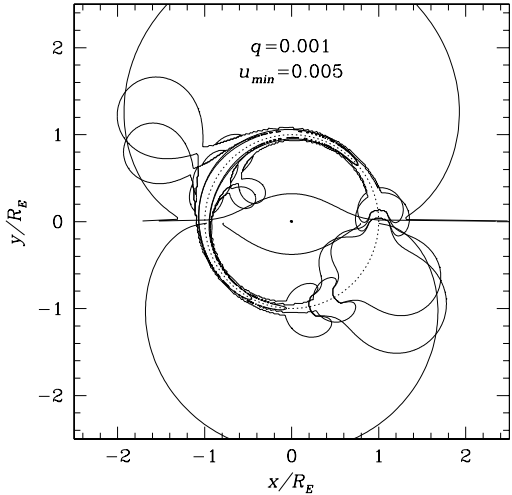}
\caption{The contours defining a $2/(S/N)$ perturbation of the
light curve grow and the area inside large curves that trace the
contour extremes along the source-lens line is increased as is the
probability for detection for smaller impact parameters $u_{min}$. The
area enclosed by the contours that is beyond the traces of the
extremes means we underestimate the true probability of detection by a
few percent for the smallest impact parameters. Here the fractional
perturbation is fixed at 0.1.
\label{fig:contoursweep2}}   
\end{figure}

The details of this averaging process are given in Appendix C.
In the average represented by Eq. (\ref{eq:fxpave}), $a$ and $\zeta$
are fixed, and the average is performed as a function of $x_p$. The
Holtzman {\it et al.} (1998) MF, which we represent as being $\propto 
M^{-1}$ for $0.08\leq M\leq 0.7M_{\s\odot}$ and $\propto M^{-2.2}$
for $0.7\leq M\leq 2M_{\s\odot}$, is used in the averaging. 
The lens distribution is determined by the triaxial Galactic bulge model
of Zhao (1996) combined with the double exponential disk model of
Bahcall and Soneira (1980).  

\begin{figure}[ht]
\plotone{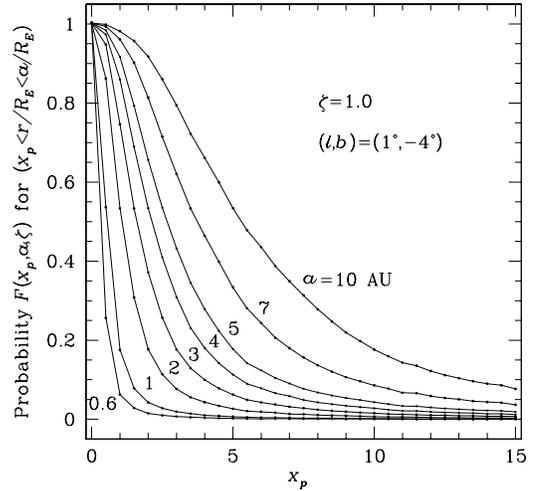}
\caption{Cumulative probability $F(x_p,a,\zeta)$ that the
projected position of the planet lies between $x_p$ and $a/R_{\s E}$
for several values of semimajor axes $a$ for a fixed source distance
$\zeta=1$.  The probability is averaged over the line of sight
distribution of lenses according to the Zhao bulge model and the
Bahcall and Soneira disk model, and over the Holtzman {\it et al.}
mass function. \label{fig:cumprobvssem}}   
\end{figure}

Fig. \ref{fig:cumprobvssem} shows the averaged cumulative probability
$F(x_p,a,\zeta)$ of the projected separation of the planet from
its star for several values of the orbit semimajor axis $a$ for the
particular choice of a LOS toward Baade's window
$(\ell,b)=(1^\circ,-4^\circ)$ ($\ell,\,b$ are Galactic longitude and
latitude respectively) and for a source distance of 8 kpc
($\zeta=1$).  The Holtzman {\it et al.} (1998) MF is
adopted for both disk and bulge stars.  Of course the cumulative 
probability is unity for $x_p=0$, since it is certain that
the projected position of the planet will be somewhere between 0 and
$a$.  The only contributions to $F(x_p,a,\zeta)$ come
from those lenses along the LOS where $R_{\s E}<a/x_p$ (Appendix
C). For small $a$, the number of lenses whose $R_{\s E}$ values
qualify plummets rapidly as $x_p$ increases. Increasingly small $M$
with $z$ either very small or very near $\zeta$ are required until
there are no lenses at all along the LOS that satisfy the condition.
$F(x_p,a,\zeta)$ in Fig. \ref{fig:cumprobvssem} correspondingly
decreases most rapidly toward zero when $x_p$ increases for small $a$
and less rapidly for larger $a$. 

\begin{figure}[ht]
\plotone{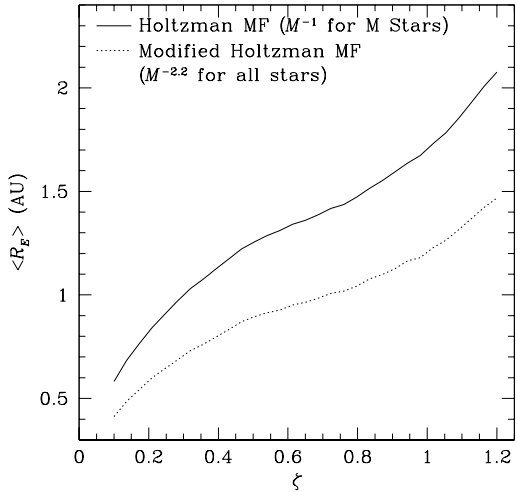}
\caption{The Einstein ring radius averaged over the lens density
distribution along the line of sight toward Baade's window
($\ell,b)=(1^\circ,-4^\circ)$ and over the indicated mass
functions for the Zhao (1996) bulge and Bahcall-Soneira (1980) disk models.
\label{fig:reavefxddos}} 
\end{figure}

Useful in understanding Fig. \ref{fig:cumprobvssem},
Fig. \ref{fig:reavefxddos} shows the Einstein ring radius averaged 
along the LOS to distance $\zeta$ according to  
\begin{displaymath}
\langle R_{\s E}\rangle=(8.07\,{\rm AU})\times
\end{displaymath}
\begin{equation}
\times\frac{\displaystyle
\int_{M_{min}}^{M_{max}}\int_0^\zeta\sqrt{\frac{Mz(\zeta-z)}
{\zeta}}z^2\frac{dn_{\s L}(M,z)}{dM}dzdM}
{\displaystyle\int_{M_{min}}^{M_{max}}\int_0^\zeta z^2\frac{dn_{\s
L}(M,z)}{dM}dzdM}, \label{eq:reavefxddos}
\end{equation}
where $dn(M,z)/dM$ (given in Appendix C) is the number density of
lenses per unit mass 
interval at position $z\leq\zeta$ along the LOS. The integrals are
evaluated with a Monte Carlo technique (Press {\it et al.} 1986).
$\langle R_{\s E}\rangle$ is shown for two MFs in
Fig. \ref{fig:reavefxddos}, but concentrate on the one for the
Holtzman {\it et al.} mass function for now.  The relatively small
value of $\langle R_{\s E}\rangle$ for $\zeta=1$ results from the
facts that the MF is dominated by M stars with small $R_{\s E}$ and
that $R_{\s E}$ decreases as the lens approaches the source at
$\zeta=1$ where the number density is highest. The curve for the
modified Holtzman {\it et al.} MF ($\propto M^{-2.2}$ in
Fig. \ref{fig:reavefxddos}) will be used later to illustrate the effect
of changing the MF on the detection probability. 
The probability density of
the planet separation is the negative derivative of these curves so
the planet has the highest probability of being found at a value of
$x_p$ where the slopes are steepest. The curves
approach zero slope more rapidly for the smaller semimajor axes, since
it is less likely to find a close planet much outside the Einstein
ring radius for the given MF.  Similarly, the planet is more likely to
have a large value of $x_p$ if the semimajor axis is larger, so the
value of $x_p$ where the slope is steepest for a given $a$ increases
with $a$. The magnitudes of the slopes decrease as $a$ increases since the
probability density is spread over a greater area and must be thereby
smaller. The probability density of the projected position is sharply
peaked near $r=a$ (Gould and Loeb, 1992).  This is understood
intuitively since the planet spends a higher fraction of the time near
the extremes of its projected orbit as it travels toward or away from the
observer. Or from the uniform distribution over the surface of the
sphere, a large area of the sphere is covered by small changes in the
LOS near the extreme.  The steepest slope for, say, $a=2$ AU in
Fig. \ref{fig:cumprobvssem} occurs near $x_p=1$, which reflects the
fact that the average $R_{\s E}$ over the LOS to $\zeta=1$
is about 1.7 AU. (That is, $x_p=r/R_{\s E}$ and $r$ prefers to be
near $a$ because of the peak at $a$ in the probability density.)
The steepest slope for $a=10$ AU occurs near $x_p=5$ or 6, which is
again the planet preferring to be at an extreme separation with
$\langle R_{\s E}\rangle\approx 1.7$ AU.  

\begin{figure}[ht]
\plotone{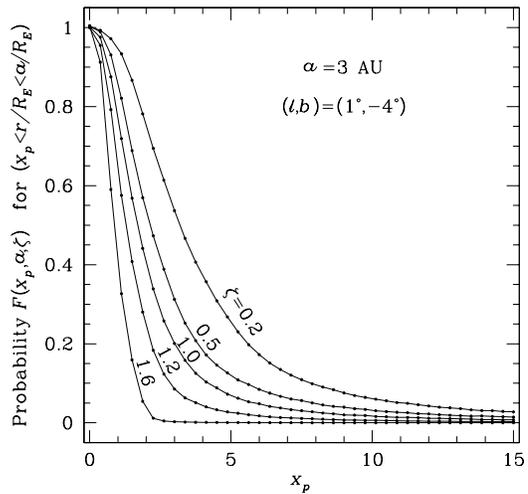}
\caption{Same as Fig. \ref{fig:cumprobvssem} except the cumulative
probability is given  for several values of source distance $\zeta$
for fixed semimajor axis $a=3$ AU. \label{fig:cumprobvszeta}}
\end{figure}

Fig. \ref{fig:cumprobvszeta} shows the same cumulative probability
distribution as Fig. \ref{fig:cumprobvssem}, but now $a$ is fixed at 3
AU and the distance to the source is varied. Again contributions to
$F(x_p,a,\zeta)$ come only from those lenses with $R_{\s E}<a/x_p$.
From Fig. \ref{fig:reavefxddos} we see that $\langle R_{\s
E}\rangle$ is large for large $\zeta$, so as $x_p$ increases, the
number of lenses along the LOS contributing to $F(x_p,a,\zeta)$
decreases most rapidly for large $\zeta$. That 
is, if $\langle R_{\s E}\rangle$ is large, the number of lenses with
sufficiently small $R_{\s E}$ to contribute to $F(x_p,a,\zeta)$
plummets most rapidly as $x_p$ increases for large $\zeta$, and 
the number decreases less rapidly for smaller $\zeta$ when $x_p$
increases. These trends are evident in Fig. \ref{fig:cumprobvszeta}.       

Since the probability density of the projected planet position peaks
near $r=a$, the planet is most likely to be separated from the lens by
$\sim 3$ AU for all the curves in Fig. \ref{fig:cumprobvszeta}. 
The steepest part of the curves indicating the highest probability
density  should then occur near where $x_p=3 {\rm AU}/\langle R_{\s
E}\rangle$. From Fig. \ref{fig:reavefxddos}, $\langle R_{\s E}\rangle
\approx 0.8,\,2.05$ AU at $\zeta=0.2,\,1.2$ respectively, and we see
that the steepest slopes indeed occur near $x_p=3\approx$ 3AU/0.8AU for
$\zeta=0.2$ and near $x_p=1.5\approx$ 3AU/2.05AU for $\zeta=1.2$. 
Another way to look at the behavior of $F(x_p,a,\zeta)$ in
Fig. \ref{fig:cumprobvszeta} is to notice that for small $\zeta$,
$\langle R_{\s E}\rangle$ is small so there is a  greater probability
of finding the planet at larger values 
of $x_p$. The consistencies in the positions of the  curves as a
function of the parameters in both Figs. \ref{fig:cumprobvssem} and
\ref{fig:cumprobvszeta} give one some confidence that the cumulative
probabilities are correctly averaged. 

The following describes how the integral of the probability density
over the area swept in the lens plane out by the contours of $2/(S/N)$
fractional perturbation of the light curves is carried out using
$F(x_p,a,\zeta)$. This is the integral of $dF(x_p,a,\zeta)/dx_p$ in
Eq. (\ref{eq:grandint}).
Assume for the present that the normalized distances from the lens to
the inner and outer traces of the amplification contour extremes in 
upper half plane of Fig. \ref{fig:contoursweepiband} are $r_1/R_{\s E}$
and $r_2/R_{\s E}$ 
respectively.  Since $F(x_p,a,\zeta)$ is the averaged cumulative probability
of finding the projected position of the planet in the annular region
between $x_p$ and $a/R_{\s E}$ centered on the lens, the probability
of finding the planet in a narrow wedge of angular width
$\Delta\theta$ and lying between $r_1/R_{\s E}$ and $r_2/R_{\s E}$ is
just $[F(x_p=r_1/R_{\s E},a,\zeta)-F(x_p=r_2/R_{\s
E},a,\zeta)]\Delta\theta/2\pi$. The 
probability of finding the planet in the area inside the two trace
curves is then the sum of these probabilities for all the positions of
the source as it is stepped through the encounter.  A similar
procedure determines the probability that the projected position of
the planet lies in the area enclosed by the trace curves in the lower
half plane of Fig \ref{fig:contoursweepiband}.  The sum of these two
probabilities is the probability of detecting the planet during this
particular encounter.  This probability is of course dependent on the
estimate of the $S/N$ (Section \ref{sc:sn}) we use to define the
perturbation contours of detectability in, for example,
Figs. \ref{fig:contoursweepiband}, \ref{fig:extrcrvsmagcmpreiband} and
\ref{fig:contoursweep2}.  

\section{Procedure \label{sc:procedure}}
For each impact parameter, the source is located by its angular
position $\theta$ measured from the point of closest approach to the
lens, and the source is stepped along its trajectory in equal increments
$\Delta\theta$.  The first point is the first encounter of the source with
the Einstein ring of the lens where the unperturbed amplification
$A_0= 1.34$, which is here assumed to be the minimum amplification for an
alert to the event.  The brightness of the source is increased by the
instantaneous amplification to determine the $S/N$ at that position.
The  extremes in the contours corresponding to the perturbation of
$2/(S/N)$  are calculated for that $S/N$ as outlined
above and in Appendix B, and the probability that the
planet lies in the wedges of angular width $\Delta\theta$ along the
source-lens line and between the extreme curves allowing detection is 
calculated for a particular semimajor axis.  These ``wedge''
probabilities are summed for the 
encounter out to a source position along its trajectory of $2R_{\s E}$
beyond its point of closest approach. The detection probability is
determined for each of a distribution of approximately thirty impact
parameters for the given source distance and magnitude, and averaged
over the impact parameters.

\begin{figure}[ht]
\plotone{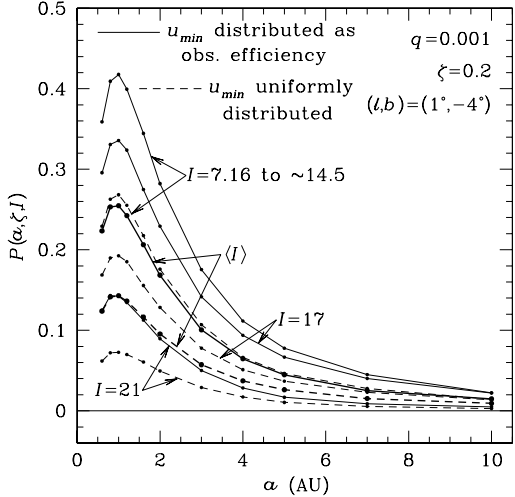}
\caption{Probability $P(a,\zeta,I)$ (thin lines) of detecting a planet
when the source at
$\zeta=0.2$ for a selection of source magnitudes plus the average of
the probabilities over the LF $P(a,\zeta)$ (heavy lines). $I=7.16$ is
the brightest star in the
LF for $A_{\s I8}=0.76$, and a single curve applies for $I=7.16$ to
about 14.5 because of our constraint that $S/N$ not exceed
200. The probability has been averaged over the LOS distribution of
lenses, over the Holtzman {\it et al.} mass function, and over the
distribution of impact parameters $u_{min}$, where the
Zhao bulge model and the Bahcall-Soneira disk model are assumed for
the stellar mass distribution in the Galaxy. \label{fig:detprobzeta.2}}
\end{figure}

\begin{figure}[ht]
\plotone{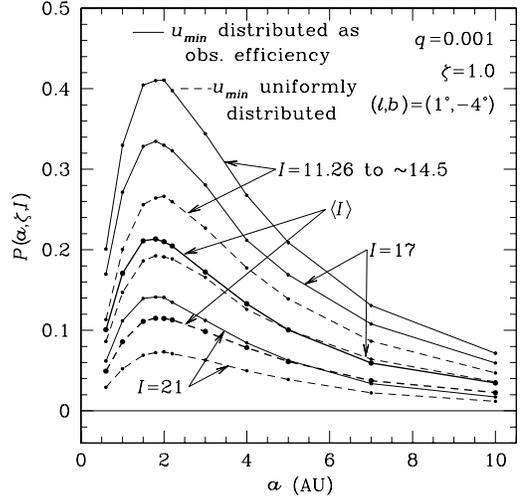}
\caption{Same as Fig. \ref{fig:detprobzeta.2} except the source is
at $\zeta=1.0$ and the brightest star in the LF is
$I=11.26$. \label{fig:detprobzeta1}}
\end{figure}

The calculation is repeated for each of 11 semimajor axes ($0.6<a<10$ 
AU), and one obtains a series of functions of $a$, labeled by the $I$
magnitude of the source at the distance $\zeta$. These are
probabilities $P(a,\zeta,I)$ of detecting the planet during the event
averaged  over the MF of the lenses, over the distribution of the
lenses along the particular LOS from the observer to the source at
$\zeta$, and over the distribution of impact parameters.
Figs. \ref{fig:detprobzeta.2} and \ref{fig:detprobzeta1} show examples
of these curves for $\zeta=0.2$ and 1 respectively for a selection of
source $I$ magnitudes.  In these figures, $q=0.001$, the LOS
is toward Baade's window $(\ell, b)=(1^\circ,-4^\circ)$ and the
extinction $A_{\s I8}=0.76$ mag (Holtzman {\it et al.} 1998).  At
$\zeta=0.2$, the brightest star in the LF has I=7.16, and the curves
for $7.16<I\ltwid 14.5$ are superposed, because for these bright stars,
$S/N$ exceeds 200 for a 60 sec integration time, and we have assumed
the integration is reduced so that $S/N$ does not exceed 200 in such
circumstances. This means that the detection probability is
observationally controlled to be the same for the given LOS and 
parameter values for all sources with $I\ltwid 14.5$.  Otherwise,
the detection probability increases with source  brightness in these
figures because the higher $S/N$ increases the area contained within
the curves tracing the extremes in the amplification contours as shown
in Fig. \ref{fig:extrcrvsmagcmpreiband}.  

We pointed out above that the probability distribution of the
projected position of the planet on the lens plane is strongly peaked
near $r=a$, and because the perturbation of the light curve decreases
monotonically as the planet is moved radially away from the Einstein
ring radius in either direction (provided it is outside the closed
critical curves), the probability of 
detection is maximized when the planet is near the Einstein ring of
the lens.  One then expects the probability of detection to be
maximized when the semimajor axis of the planet is near the average
value of the Einstein ring radius between the observer and the source
location. From Fig. \ref{fig:reavefxddos}, at $\zeta=0.2,\,1.0$,
$\langle R_{\s E}\rangle\approx 0.85,\,1.7$ AU for the Holtzman {\it
et al.} (1998) MF, where relatively small values result
from the dominance of M stars in the MF. Correspondingly, the peak in
the detection probabilities occur near (1 AU, 2 AU) for
$\zeta=(0.2,\,1.0)$

The probabilities are shown for two distributions of impact parameters
in Figs. \ref{fig:detprobzeta.2} and \ref{fig:detprobzeta1}.  The
random nature of the events means that the impact parameters will be
uniformly distributed, where we choose $0<u_{min}<1$ with $u=1$
corresponding to an amplification of 1.34 being the assumed threshold
for an alert. However, events with smaller impact
parameters are more likely to be observed because of the higher
magnifications, so the distribution of impact parameters of recorded
events will be biased toward the smaller values.  Accordingly, the
second distribution of $u_{min}$ is that of Fig. 9 from Alcock {et
al.} (2000), which has a higher proportion of small impact parameters
to reflect the observational efficiencies. In fact, events with
$u_{min}>0.9$ contribute negligibly to the average of the
probabilities over the distribution of $u_{min}$, so we choose 30
values $0.01\leq u_{min}\leq 0.91$ for the uniform distribution.  The
distribution determined by observational efficiencies naturally
reduces the number of larger impact parameters in the range
$0<u_{min}<1$. The higher probabilities for this latter
distribution of $u_{min}$ reflect the higher probabilities for closer
impact parameters implied in Figs. \ref{fig:contoursweepiband} and
\ref{fig:contoursweep2}.

\begin{figure}[ht]
\plotone{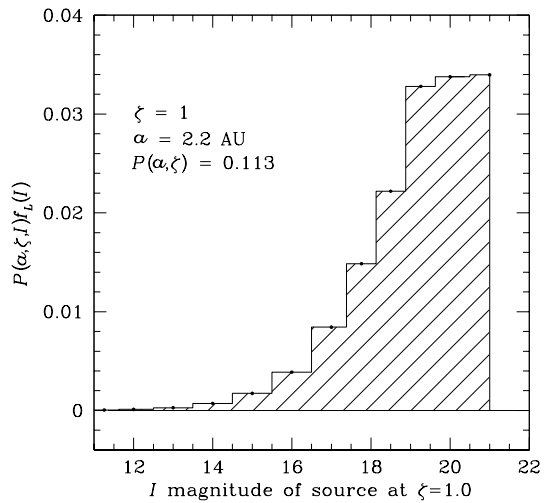}
\caption{An example of the average of the detection probabilities
over the LF for $\zeta=1$ and $a=2.2$ AU. The probabilities have
already been averaged over the the distribution of lenses along the
LOS, over the Holtzman {\it et al.} mass function and over a
uniform distribution of impact parameters. The shaded area under the
histogram is the averaged probability, which is 0.113 in this
case. \label{fig:aveoverlf}}
\end{figure}

\begin{figure}[ht]
\plotone{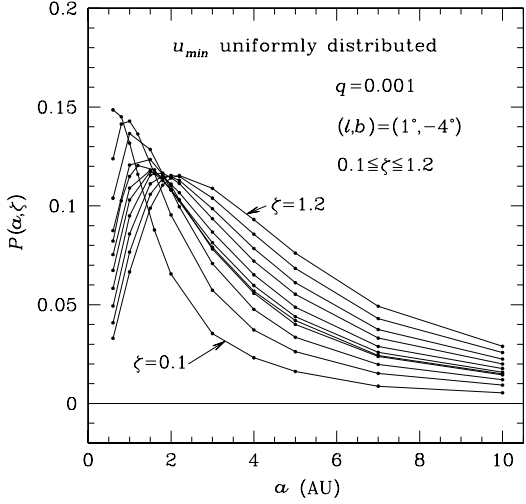}
\caption{Average of the detection probability over the LOS
distribution of lenses, over the Holtzman {\it et al.} MF, over a
uniform distribution of impact parameters $u_{min}$, and over the
LF for $\zeta$ incremented in steps of 0.1 (except $\zeta=0.4$ is not
included).  The peaks occur where $a\approx\langle R_{\s
E}\rangle$. \label{fig:pavelf}}
\end{figure}

Also shown in Figs. \ref{fig:detprobzeta.2} and \ref{fig:detprobzeta1}
are the averages over the distribution of $I$ magnitudes of the sources
at each distance.  These are effectively averages over the LF
expressed in $I$ magnitudes appropriate to each distance but truncated
to the visible sources ($I<21$).  The averaged probability is 
\begin{equation}
P(a,\zeta)=\int_{I_{min}}^{21} P(a,\zeta,I)f_{\s L}(I)dI,
\label{eq:pavelf}
\end{equation}
where $I_{min}=10.5+A_{I8}\zeta+5\log{\zeta}$ and $f_{\s
L}(I)$ is $f_{\s L}(I_8)$ in Eq. (\ref{eq:lfi}) with
Eq. (\ref{eq:i8vsiappar}) replacing $I_8$ and the result renormalized so
that $f_{\s L}(I)dI$  represents the fraction of {\it visible} sources
in range $dI$ about $I$. Examples of $P(a,\zeta,I)$ are the labeled
curves in Figs. \ref{fig:detprobzeta.2} and \ref{fig:detprobzeta1},
and $P(a,\zeta)$ is labeled with $\langle I\rangle$. 
Since the curves are for discrete values of $I$, the integral in
Eq. (\ref{eq:pavelf}) is approximated by $\sum_i
P(a,\zeta,I_i)f_{\s L}(I_i)\Delta I_i$, where $I_i$ are the discrete
values of $I$ at which $P$ is evaluated and $\Delta I_i$ is an
increment centered on $I_i$ and stretching halfway to adjacent values
of $I_i$ on either side.   Fig. \ref{fig:aveoverlf} demonstrates this
procedure, where the shaded area under the histogram is the average
over the LF for $\zeta=1$ and $a=2.2$ AU.  Fig. \ref{fig:pavelf} shows
the detection probability averaged over the luminosity function of the
sources for the uniform distribution of impact parameters, over the
Holtzman {\it et al.} MF and over the LOS 
distribution of lenses for a series of values of $\zeta$.  The
semimajor axis where the probability is maximal follows the mean value of
$R_{\s E}$ as described above.  The derivation of the overall averaged
detection probability will be completed by the average over the
distribution of visible sources. 

\section{Average of $P(a,\zeta)$ over$\zeta$ \label{sc:paveoverzeta}}
The final average of the detection probability is carried out with the
following expression:
\begin{equation}
P(a)=\frac{\displaystyle\int_{0.1}^{1.2}P(a,\zeta)
n_s(\zeta)\zeta^2\xi({\cal F}_{21},\zeta,A_{\s I8})d\zeta}{\displaystyle
\int_{0.1}^{1.2}n_s(\zeta)\zeta^2 \xi({\cal F}_{21},\zeta,A_{\s I8})d\zeta}, 
\label{eq:paveoverzeta}
\end{equation} 
where $n_s(\zeta)$ is the number density of sources at distance
$\zeta$ and where
\begin{displaymath}
\xi({\cal F}_0,\zeta,A_{\s I8})=C_1(4\pi D_8^2{\cal F}_{21}\zeta^2e^
{-0.921A_{I8}}) ^\beta+C_2, \;\;\;(C11)
\end{displaymath}
derived in Appendix A, is the fraction of the sources with observed
flux densities $\cal F$ 
greater than a minimum ${\cal F}_{21}$ assumed detectable. 
${\cal F}_{21}=6.21\times 10^{-15}$ ${\rm ergs\,cm^{-2} sec^{-1}}$ is the
minimum flux density in $I$-band considered observable---here from a
star with $I=21$. $C_1L_{\s I0}^\beta+C_2$ is the fraction of the
sources with luminosity $L_{\s I}>L_{\s I0}$, which follows from the
fraction of sources with $I<I_0$ found from the LF in Appendix
A.  The constants and index are different in the LF for
each of three ranges of $I_8$ (Eq. (\ref{eq:lfi})) with 1.16\% of the
stars having $I_8<17$ and 4.18\% of the stars having $I_8<18.5$, with
$I=17$ and 18.5 being the values where the LF index changes.  Hence,
$C_1,\,C_2$ and $\beta$ must change when $\xi=0.0116$ and
0.0418. Also $\xi=1$ for values of $\zeta$ so small
that all of the sources are visible and $\xi=0$ for values of
$\zeta$ so large that even the brightest source in the LF has $I>21$. The
parameters in $\xi$ and examples of values of $\zeta$
where changes occur for a few values of $A_{\s I8}$ are given in
Appendix A.  Fig. \ref{fig:visibilityi21} shows $\xi$ as a
function of $\zeta$ for three values of $A_{I8}$.
\begin{figure}[ht]
\plotone{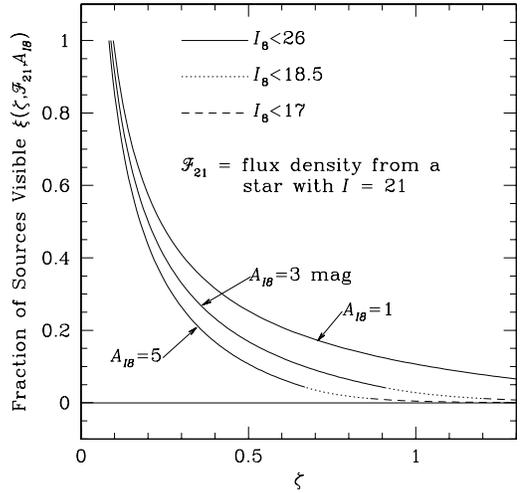}
\caption{Fraction of sources visible for the adopted LF as a function
of source distance. ${\cal F}_{21}$ is the flux density from a star
with $I$-magnitude = 21 that is assumed to be the minimum observable,
and $A_{\s I8}$ is the $I$-band extinction to 8
kpc. The different line types correspond to the different forms of the
LF for different brightness ranges. \label{fig:visibilityi21}}
\end{figure}

We choose source distances for which the detection probability is to
be determined to range from $\zeta=0.1$ to 1.2 in increments of
0.1. It is very unlikely for any source to be closer to us than
$\zeta=0.1$ (0.8 kpc), and beyond $\zeta = 1.2$, extinction and
blending will probably make the number of sources that are usably visible
negligibly small. The source number density along the LOS is
obtained from the Zhao (1996) bulge model and the Bahcall-Soneira disk
for the Holtzman {\it et al.} (1998) mass function in Appendix B.  The
overall coefficients in the bulge and disk distributions are determined
by the constraints on the total bulge mass of $2.2\times
10^{10}M_{\s\odot}$ (Zhao, 1996) and a local disk mass density of
$0.05{\rm M_{\s\odot}pc^{-3}}$ (Bahcall and Soneira, 1980) respectively.
The functions $P(a,\zeta)$ ({\it E.g.}, Fig. \ref{fig:pavelf})
are made continuous functions of $\zeta$ by interpolating between the
discrete points where they are evaluated. This makes the integrands
for a given LOS in Eq. (\ref{eq:paveoverzeta}) functions of
$\zeta$ alone and evaluation of the integrals is straightforward.  

\begin{figure}[ht]
\plotone{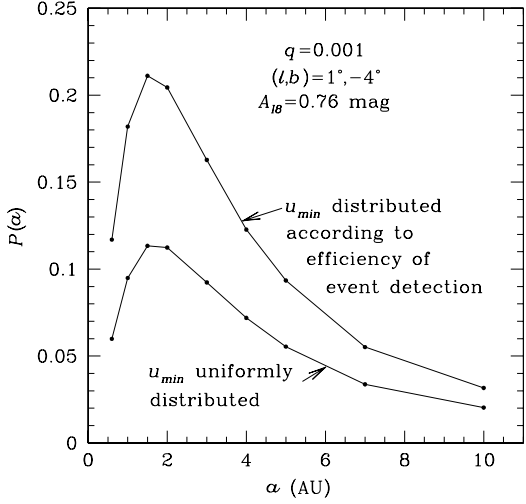}
\caption{Averaged probability of detecting a planet during a
microlensing event toward Baade's window for a uniform distribution of
impact parameters and a distribution determined by the observational
efficiency of the MACHO group (Alcock {\it et al.} 2000).
\label{fig:paveoverzeta}}
\end{figure}

Fig. \ref{fig:paveoverzeta} shows the averaged probability $P(a)$ for
a uniform distribution of impact parameters and for 
that distribution appropriate to the efficiency of detection by the
MACHO group (Alcock {\it et al.}, 2000).  The line of sight is toward
Baade's window $(\ell,b)=(1^\circ,-4^\circ)$, where $A_{\s
I8}=0.76$ (Holtzman {\it et al.} 1998). The higher weight given to 
the small impact parameters in the latter distribution nearly doubles
the probability of detection obtained from a uniform distribution of
$u_{min}$. The large probabilities at small planet separations that
resulted for small $\zeta$ (Fig. \ref{fig:pavelf}) are seen to be
suppressed by the lack of sources at small $\zeta$. The semimajor axis
for the peak probability of detection is near 1.5 AU, which is close
to the mean $R_{\s E}$ over the entire LOS for the Holtzman,
{\it et al.} mass function and Zhao plus Bahcall-Soneira galactic model.  

\section{Parameter dependence of the detection probability
\label{sc:paramdep}} 
\begin{figure}[ht]
\plotone{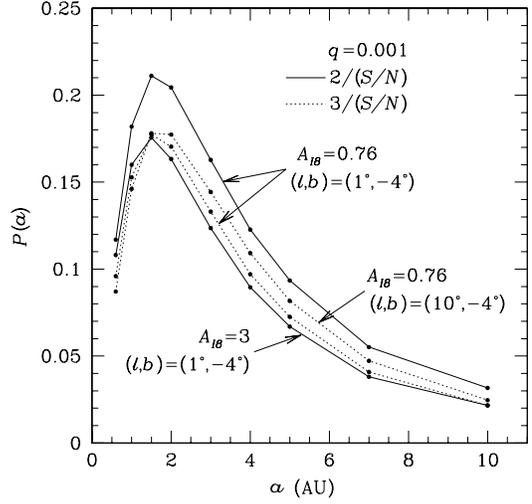}
\caption{Averaged probability of detecting a planet during a
microlensing event showing effect of shifting the LOS, of increasing
the extinction, and of using a $3/(S/N)$ instead of a $2/(S/N)$
detection criterion. The probability is averaged over the source-lens
impact parameter distribution determined by the observational
efficiency, over the lens distribution along the LOS, over the MF,
over the LF, and over the visible source distribution along the
LOS. \label{fig:paveoverzeta2}}
\end{figure}

Fig. \ref{fig:paveoverzeta2} shows the reduction in the probability by
increasing the extinction to $A_{\s I8}=3$, by assuming a $3\sigma$
instead of a $2\sigma$ detection both along the LOS to Baade's window
and by changing the LOS to $(\ell,b)=(10^\circ,-4^\circ)$ with $A_{\s
I8}$ restored to 0.76 mag.  All assume the distribution of $u_{min}$
appropriate to the efficiency of observing the event, and this
curve for Baade's window and $A_{\s I8}=0.76$ is repeated from
Fig. \ref{fig:paveoverzeta} for comparison. The $3/(S/N)$ detection
criterion reduces the probability of detection for all semimajor axes
over that for a $2/(S/N)$ detection criterion as expected, and the
mechanism is a reduction in the area in the lens plane inside of which
a planet's projected position would lead to a detection.   Changing
the line of sight to $(\ell,b)=(10^\circ,-4^\circ)$ (also with the
$3/(S/N)$ detection criterion) from ($\ell,b)=(1^\circ,-4^\circ)$ with
equal extinctions increases the probability of detection for 
$a\gtwid$ its value corresponding to the peak probability and
decreases the probability for $a\ltwid$ its value near the peak.  By
swinging the LOS away from the 
bulge, we have effectively increased the mean Einstein ring radii of the
lenses.  There are considerably fewer sources and fewer lenses at the
most distant parts of the LOS at the larger Galactic longitudes, so
there is less weight given to those configurations of small Einstein ring
radii in the averages over both the lens and source distributions.
That means that planets with small semimajor axes (near $R_{\s E}$
where the probability is large) will have a smaller probability of
detection, whereas those with larger semimajor axes will have their
averaged probabilities increased.  One can see from
Fig. \ref{fig:paveoverzeta2}  that this is not a large effect. 
The increase in $A_{\s I8}$ toward Baade's window has the effect of
increasing the apparent magnitudes of all the sources, thereby
reducing the $S/N$ and the probability of detecting a planet. 

\begin{figure}[ht]
\plotone{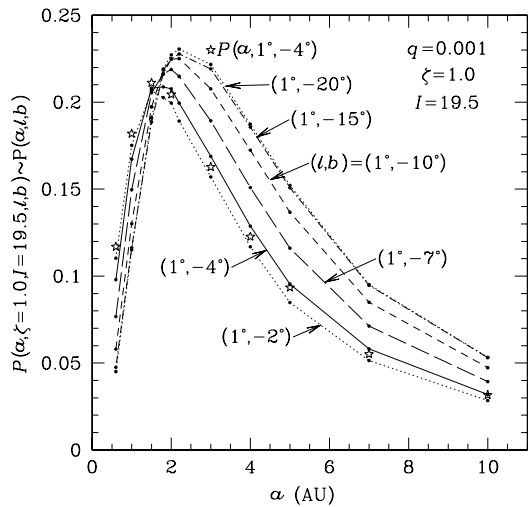}
\caption{Averaged probability of detecting a planet during a
microlensing event showing the effect of shifting the line of sight
along a line of constant galactic longitude $\ell=1^\circ$.  Here
$P(a,\zeta,I)$, with $\zeta=1$ and $I=19.5$, is used to approximate
$P(a)$, where the galactic coordinates have been added in the
arguments to distinguish the LOSs.  The stars correspond to
$P(a)\equiv P(a,1^\circ,-4^\circ)$ and indicate the  good
approximation by the partial average for the selected $I$ magnitude
for$(\ell,b)=(1^\circ,-4^\circ)$.
\label{fig:loscmpre}}
\end{figure}

Fig. \ref{fig:loscmpre} emphasizes the effect of changing the line of
sight relative to the center of the Galaxy. Here a series of LOS with
different Galactic latitudes and the same Galactic longitude
($\ell=1^\circ$) again show the effect of decreasing or increasing the
proportion of the lenses for large distances from the observer.  The
probability of detection is raised for larger semimajor axes for the LOS
directed further from the high spatial density region near the
Galactic center, and it is lowered for the LOS closer to the
Galactic center for reasons discussed above.  Rather
than show the complete averaged probability for these examples, we notice
that the probability averaged only over the distribution of impact
parameters, over the distribution of lenses along the line of sight
to $\zeta=1.0$ and over the MF for $I=19.5$ mag gives a 
good approximation to the detection probability with additional averages
over the LF and the distributions of sources leading to $P(a)$.  The
points indicated by stars in 
Fig. \ref{fig:loscmpre} are appropriate to $P(a)$ and indicate how
well the partial average for the 
selected values of $\zeta$ and $I$ approximates the total average for
$(\ell,b)=(1^\circ,-4^\circ)$.  

The peak probability, in
addition to being shifted slightly to larger semimajor axes as $|b|$ is
increased, is also increased slightly.  This is an extension of the
increase in probability for larger semimajor axes as less weight is
given to the more distant lenses with increasing $|b|$. Since the
averaged probability depends in an intricate way on the weighting of
the Einstein ring radius by the distribution of lenses along the LOS 
relative to the semimajor axis of the planetary orbit, there 
is no reason to expect the peak probability not to change as it shifts
to larger semimajor axes. There is almost no difference in the
probability in Fig. \ref{fig:loscmpre} for $b=-15^\circ$ and
$-20^\circ$. The LOS for these latitudes does not intercept much of
the bulge distribution of stars, so the distribution of lenses along
the LOS, being mostly disk stars, changes very little between
$b=15^\circ$ and $20^\circ$.

\begin{figure}[ht]
\plotone{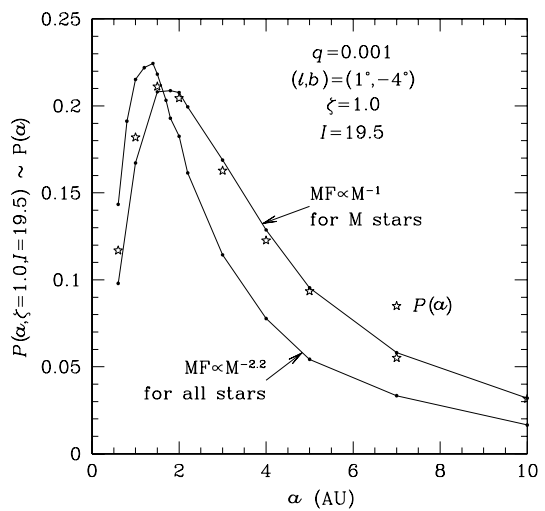}
\caption{Effect on the averaged detection probability of changing the
MF. The other information is the same as in Fig. \ref{fig:loscmpre}.
The larger number of M stars in the MF with the larger index leads to
a shift in the peak probability to smaller $a$ because of the smaller
$\langle R_{\s E}\rangle$. This same shift in $\langle R_{\s E}$ to
smaller values leads to a decreased probability for large $a$ and an
increased probability for small $a$. \label{fig:pavemfcmpre}}
\end{figure}

The reason for considering the modified Holtzman {\it et al.} mass
function in Fig. \ref{fig:reavefxddos} is that more M stars than
are observed seem to be required in the MF to account for the relatively
large number of short time scale events toward the Galactic center.
The modified Holtzman mass function $\propto M^{-2.2}$ for
all stars $0.08\leq M\leq 2.0M_{\s\odot}$ gives a reasonably good
approximation to the time scale frequency distribution obtained by the
MACHO group for the 1993 bulge season (Peale, 1998), but extending the
MF into the brown dwarf region does not (Peale, 1999). The larger number
of small mass stars in this distribution leads to the reduction in the
averaged Einstein ring radius for all $\zeta$ in Fig.
\ref{fig:reavefxddos}.  Its consideration here allows us the
opportunity to illustrate the effect of changing the mass function on
the probability of planet detection.   We expect this reduction in the
mean Einstein ring radius to shift the peak in the probability
distribution to smaller semimajor axes, to lower the probability
for the larger semimajor axes, and to raise it for the smallest semimajor
axes as discussed above. This expected effect is shown in 
Fig. \ref{fig:pavemfcmpre}, where we have again used the partial
average of the detection probability as an approximation to $P(a)$.
The peak probability has moved from something near 1.8 AU 
to being near 1.4 AU, again showing that the peak occurs near the
averaged Einstein ring radius $\langle R_{\s E}\rangle\approx 1.4$ for
the largest values of $\zeta$ in Fig. \ref{fig:reavefxddos} (lower
curve).  

\begin{figure}[ht]
\plotone{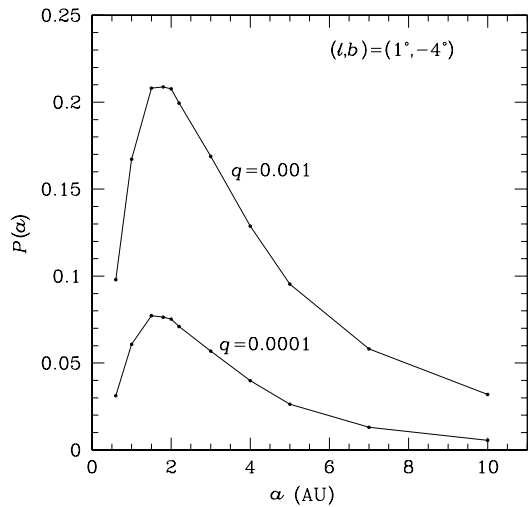}
\caption{Effect on the completely averaged detection probability of
changing the planet-star mass ratio $q$ showing that
$P(a)\sim\sqrt{q}$.  The actual ratio between the two curves is 3.14
at $a=0.6$, decreases to 2.7 near the peak and increases to 5.7 at
$a=10.0$, so the scaling is poorest at the larger values of $a$.
\label{fig:moverMcmpre}}
\end{figure}

The averaged probability of planet detection varies approximately as
$\sqrt{q}$. This is shown in Fig. \ref{fig:moverMcmpre}, where
decreasing $q$ by a factor of 10 leads to a reduction in the
probability of about $\sqrt{10}$.  The actual ratio of the
probabilities for the two values of $q$ starts at 3.14 for $a=0.6$,
drops to 2.7 near the peak and rises to 5.7 for $a=10$, which shows
that the approximate scaling is rather crude for some values of
$a$, but it still gives a useful extension of the results for
$q=0.001$ and 0.0001 shown in Fig. \ref{fig:moverMcmpre}. This scaling
should not be  
carried to mass ratios less than $10^{-4}$, since effects of the
non-zero size of the source become important and are not included in
this analysis ({\it e.g.}, Bennett and Rhie, 1996). On the other hand,
the probabilities so obtained may not be  that far from the Bennett
and Rhie values for small planets (Peale, 1997). The source of this
scaling can be traced to the scaling of the size 
of the contours of equal perturbation in Figs. \ref{fig:contoursweepiband}
and \ref{fig:contoursweep2} approximately as $\sqrt{q}$ (Gould and Loeb, 1992).
The length of the arc described by the center of the contours during
the event is independent of $q$. So only the dimension of the total
area swept out by the contours that is perpendicular to this arc is
affected by $q$. As this dimension scales approximately as $\sqrt{q}$, the
whole area swept out scales approximately as $\sqrt{q}$. Since the
probability per unit area of the projected position of the planet
falling into the area usually does not vary much over the area, the
total probability of finding the planet in the area scales roughly as
the area and hence $\sim\sqrt{q}$. 

\begin{figure}[ht]
\plotone{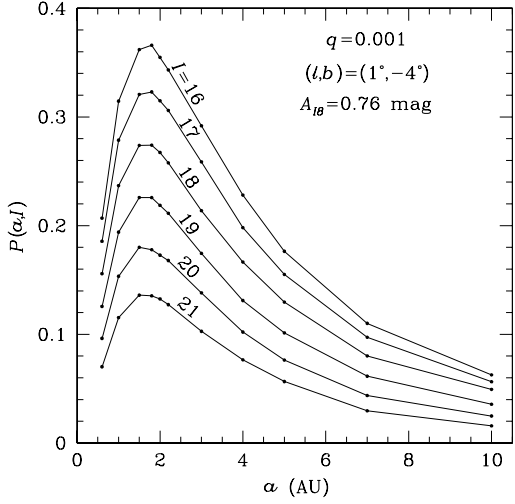}
\caption{Averaged detection probability as a function of the $I$
magnitude of the source for the LOS toward Baade's window. The average
over the LF is bypassed here and $P(a)\rightarrow P(a,I)$ on the
vertical axis. The impact parameter $u_{min}$ is distributed according
to the efficiency of event detection.  The Holtzman {et al.} (1998)
mass function is assumed. \label{fig:paveoverzetal}}
\end{figure}

We have averaged over the LF in the above probabilities, but since the
magnitude of the source is known in any particular event, we could
eliminate this average and evaluate the detection probability for
particular $I$ magnitudes.  These probabilities are shown in
Fig. \ref{fig:paveoverzetal}.  The
probabilities for fixed $I$ magnitudes averaged over the distribution of
lenses along the LOS, over the lens MF, over the distribution of impact
parameters (here determined by observational efficiency), and over the
distribution of visible sources along the LOS  are higher for
brighter stars because of the increasing $S/N$ with decreasing $I$-band
magnitude. As
in the other examples, the probabilities are significant even for the
faintest stars that we have assumed suitable for followup photometry
from the ground.

\begin{figure}[ht]
\plotone{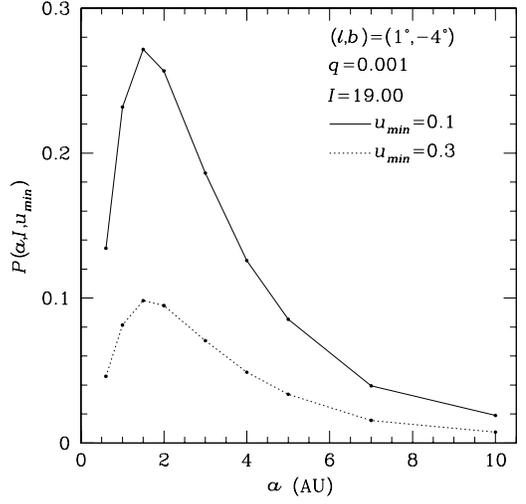}
\caption{Probability of detecting a planet during an individual event
where both the impact parameter and the $I$ magnitude of the source are
known. The averages over the impact parameter distribution and over
the LF are omitted, leaving only averages over the degenerate
parameters, $z,\,\zeta$ and $M$, where the former two are
weighted by the distribution of lenses and sources along the LOS and
the last by the MF.  $P(a)\rightarrow P(a,I,u_{min})$ on the vertical
axis.   Note the rapid drop in the detection probability
as the impact parameter is increased. \label{fig:indivevent}}
\end{figure}

The peak amplification during an event will yield the impact parameter
with an accuracy 
determined by how well the blending can be modeled, so the average
over $u_{min}$ can also be eliminated for individual events.
Fig. \ref{fig:indivevent} shows the detection probabilities for
individual events involving a 19th magnitude source with impact
parameters $u_{min}=0.1$ and 0.3.  The probabilities are now averaged only
over the degenerate parameters of the distances to the lens and
source, weighted by the distribution of lenses and sources along the LOS,
and the MF. The rapid drop in the detection probability as
the impact parameter is increased is evident, and it again illustrates
how the bias toward small impact parameters in the observational
record increases the completely averaged probability of detection in
Fig. \ref{fig:paveoverzeta}.

Since Fig. \ref{fig:indivevent} is based on idealized observational
conditions, it is probably of limited use in interpreting real microlensing
data from an individual event.  However, the techniques developed here
can be used in the interpretation of the Gaudi and Sackett (2000)
analysis of null results from real data (Albrow {\it et al.} 2000b) by
eliminating the guesses at representative source and lens distances.
The $\chi^2$ analysis of Gaudi and Sackett outlined in
Section 1 yields a best estimate of the impact parameter and source
$I$ magnitude, which we shall assume as known for the probability
$\epsilon(x_p,q)$ in Eq. (\ref{eq:epsbq}).
Like Fig \ref{fig:indivevent}, this eliminates averages over $u_{min}$
and over the LF of the sources in Eq. (\ref{eq:grandint}).  The
elimination of the LF average means the integration over the
distribution of sources can be included with the integration over the
distribution of lenses and over the MF.  The point where $q$ enters
the averaging process is in the integration of $dF(x_p,a,\zeta)/dx_p$
in Eq. (\ref{eq:grandint}) over the area $S(I,u_{min}^i)$ inside of
which a planet would be detected, since
$q,I,u_{min}$ determine the extent of the area.  
 To determine the probability of detecting a
planet with a particular mass ratio q at semimajor axis a for the
real data from a particular event, $\epsilon(x_p,q)$ is integrated
over the probability density $-dF(x_p,a)/dx_p$ over the range of $x_p$
where the density is non-negligible and for which $\epsilon(x_p,q)$
has been determined from the data.  Note that $\zeta$ is missing from
the arguments of $F(x_p,a)$ since we have also integrated over the
distribution of source distances weighted by the source distribution.
Hence,
\begin{displaymath}
P^{\s\prime}(a,q)=\int_{{x_p}_{min}}^{{x_p}_{max}}dx_p\epsilon(x_p,q)
\left(-\frac{d}{dx_p}\right)
\end{displaymath}
\begin{displaymath}
\times \int_{0.1}^{1.2}
\int_0^\zeta\int_{M_{min}}^{M_{max}}n_{\s S}^{\s\prime}(\zeta)
n_{\s L}^{\s\prime}(z,M)
\end{displaymath}
\begin{equation}
\times\Theta[f(x_p,a,\zeta,z,M)]F(x_p,a,\zeta,z,M)
d\zeta,dz\,dM, \label{eq:grandintdata}
\end{equation}
is the probability of detecting a planet with mass ratio $q$ at
semimajor axis $a$ for the particular event whose data were used in
the $\chi^2$ analysis to yield $\epsilon(x_p,q)$. The variables in Eq.
(\ref{eq:grandintdata}) are defined just below Eq. (\ref{eq:grandint}).
The integral is 
repeated for a sufficient number of values of $a$ to define the
functional form of $P^{\s\prime}(a,q)$. The integrals are to be
evaluated with a Monte Carlo technique as before. It should be noted
that $\epsilon(x_p,q)$ depends only on the data and is model
independent, whereas $P^{\s\prime}(a,q)$ depends on MF, and the lens
and source distances weighted by the lens and source distributions
along the LOS, 
and finally $P(a,q=0.001)$ in Eq. (\ref{eq:paveoverzeta}) depends on all
of these parameters plus the LF of the sources and the distribution of
impact parameters.
\section{Summary \label{sc:summary}}
We have derived the probability of detecting a planetary companion of
a star acting as a gravitational lens of a more distant star toward
the Galactic bulge sometime during the microlensing event.  The
probability  is determined as a function of the orbital semimajor axis
of the planet, and it is based on a criterion which assumes that a
perturbation to the microlensing light curve due to the presence of
the planetary companion that exceeds $2/(S/N)$ for at least 20
consecutive photometric data points 
is detectable.  By the observer's choice under the assumption that
many events are being monitored simultaneously and the likely
dominance of systematic noise for brighter stars, integration times are
assumed to be terminated when $S/N$ exceeds 200. Given this
criterion, the probability for $q=0.001$ (or 0.0001) is
averaged over the distribution of the projected position of the planet
relative to the lens, over the spatial distribution of the lenses
along the LOS for a model of the Galaxy incorporating the Zhao (1996)
and the Bahcall-Soneira (1980) bulge and disk model respectively, over
the Holtzman {\it et al.}  (1998) mass function, over the distribution
of source-lens impact parameters inside the Einstein ring radius
$R_{\s E}$, over the $I$-band LF derived from the $J$-band LF of
Zocalli {\it et al.} (2000) as adjusted by the source distance and the
extinction thereto, and over the spatial distribution of 
sources along the LOS, which is identical to the distribution of
lenses except only a fraction are visible due to distance and
extinction.  

The averaged probability scales approximately as $\sqrt{q}$
(Fig. \ref{fig:moverMcmpre}), but the scaling 
should not be extended to values of $q\ltwid 10^{-4}$ where the
effect of the non-zero size of the sources becomes important.  The  
detection probability has its maximum value of a little over 10\% if
a uniform distribution of impact parameters is assumed and over 20\%
if the distribution of impact parameters is adjusted for the observing
efficiency found by Alcock {\it et al.} (2000)
(Fig. \ref{fig:paveoverzeta}). The maximum occurs for
a planet semimajor axis that is close to the average value of $R_{\s
E}$ over the LOS---1.5 to 2 AU for a LOS toward Baade's window $(\ell,
b)=(1^\circ, -4^\circ)$ for the Holtzman {\it et al.} (1998) MF.  The
averaged probability of detection is decreased with increasing
extinction, because of the lowered $S/N$ for all sources.
Requiring a detectable perturbation of the light curve to be $3/(S/N)$
rather than $2/(S/N)$ decreases the probability by about 20\%
(Fig. \ref{fig:paveoverzeta2}). If the   
probability is not averaged over the LF, the otherwise averaged
probability increases with source brightness because of the higher
$S/N$ (Fig. \ref{fig:paveoverzetal}).  A MF with a higher proportion
of M stars shifts the peak in the probability to smaller semimajor
axes as expected (Fig. \ref{fig:pavemfcmpre}).  It also reduces the
probability of detecting a 
planet at large semimajor axes while increasing the probability of
detecting planets at very small semimajor axes.  These trends are a
consequence of the reduced $\langle R_{\s E}\rangle$, where preference
for semimajor axes comparable to $R_{\s E}$ result from
highest probability of detecting a planet occurring when its projected
position is near $R_{\s E}$.  These same trends in the probability are
seen when the LOS is moved away or toward regions of high source
density near the Galactic bulge (Figs. \ref{fig:loscmpre} and
\ref{fig:paveoverzeta2}), where the cause is again the change in
$\langle R_{\s E}\rangle$ because of the different weightings of the
source and lens distributions in the averages for different LOSs.

For individual events, good estimates of the impact parameter and the
magnitude of the source are usually available.  The averages over the
impact parameter distribution and over the LF are then omitted in
determining the probability of detection for a particular event
leaving averages only over the degenerate parameters.  The drop in
detection probability with impact parameter for a 19th magnitude
source is illustrated in (Fig. \ref{fig:indivevent}).  For real
individual events, the empirically determined, model independent
probability $\epsilon(x_p,q)$ is averaged over probability density
$-dF(x_p,a)/dx_p$ to yield $P^{\s\prime}(a,q)$, the probability of
detecting a planet with given planet-star mass ratio $q$ at semimajor
axis $a$ for that particular event.

\section{Discussion \label{sc:discussion}}
All of the variations of parameters we have investigated maintain the
relatively high detection probabilities over a substantial range of
semimajor axes, while shifting the peak slightly and altering the
relative probabilities of detecting planets with large and small
semimajor axes. This relative robustness reinforces the optimism for
planet detection through microlensing generated by the Gould and Loeb
(1992) paper. Continuous coverage of the light curve is assumed, whereas
weather and hardware maintenance limit the Galactic bulge coverage by
even a global 
distribution of 3 or 4 telescopes to a maximum near 80\% at the
midpoint of the bulge season, and this falls off on either side of mid
season because a smaller fraction of the night is usable (Peale,
1997).  Still, for Jupiter and Saturn mass planets, defined in terms of
the planet-star mass ratio, the perturbation will last on the order of
one day and would usually be detected, although the light curve during
the perturbation would not be completely covered.  If breaks in the
coverage exceed one or two days at a time, the probability of planet
detection will be reduced by the percentage of total event time that is
missed.  Our detection criterion for a perturbation of magnitude
$2/(S/N)$ persisting for at least 20 consecutive data points means
that missed coverage during the perturbation is serious for detection
and much more so for determining $q$.

Necessarily, we have simplified the calculation of $S/N$
for unknown observing systems.  The $S/N$ from photon statistics and sky
background is reduced by a factor 1/4 to account for systematic and
other neglected noise sources. The factor was chosen to yield the
empirical $S/N$ values for two real systems with their observing
conditions. Although this yields reasonable $S/N$ estimates for our
purposes, each system must be separately calibrated, and this constrains
other parameters such as integration time for desired $S/N\approx 100$
or more.  Increased integration time may compromise the desired
sampling time of a data point every one or two hours if there are
too many ongoing events that must be monitored.  Either the overall
probability of planet detection would be thereby reduced, or the
number of events that can be adequately followed with high $S/N$ will
be limited in a ground based search.  The averaged detection
probabilities thus explicitly depend on our assumptions of a 2 m
telescope, 1.4 arcsec seeing, and a fixed integration time of 60
seconds, because of their determination of $S/N$ for a given magnitude
source.

We have neglected blended light under the assumption that image
subtraction techniques (Alcock {\it et al.} 2000) can eliminate this
contamination for a wide range of source magnitudes and lensing
magnifications.  If unmodeled blended light remains, it will have the
effect of decreasing the fractional perturbation of the light curve by
the planet and reducing the probability of detection in addition to
introducing an uncertainty in the impact parameter.
Several parameters, such as integration time, are not going to be
fixed as we have assumed, but will vary even during a single night as
conditions change.  We have also assumed the single lens light curve
is known a priori, when only a light curve that is the best fit to the
data will be available.

The complete averages of the planet detection probabilities we have
derived, in addition to establishing the robustness of their relatively
large values, are useful  in determining the effect of various
distributions of parameter values.  For establishing the meaning of
null results, however, all the information available for each
particular event should be used.  Averaged values should be used only
for those properties of the lens-source system that cannot be lifted
from the degeneracies.  The magnitude of the source is going to be
more or less known, so averages over the LF are not appropriate in
analyzing individual events. The
peak amplification will yield the impact parameter with an accuracy
determined by how well the blending can be modeled, so the average
over $u_{min}$ should also be eliminated for individual events.   We
have shown how the probability varies for various magnitude sources,
but averaged over all other parameters in
Fig. \ref{fig:paveoverzetal}, and Fig. \ref{fig:indivevent} shows
how the probability decreases for individual events as the impact
parameter is increased. This decrease can be also be estimated from 
Fig. \ref{fig:probvsubiband}. The degeneracies remaining are the
distances to the lens and source and the mass of the lens, so for most
events, averages over the distribution of lenses and sources along the
LOS and over the MF are still appropriate. One or more of these
averages might be removed for individual events if the parallax of the
lens can be determined.  

There have been several comparisons of the microlensing technique for
the detection of planets with the very successful radial velocity
technique ({\it e.g.}, Albrow {\it et al.} 2000b), where the emphasis
has been on their complementary detection of distant and close planets
respectively or microlensing's unique sensitivity to terrestrial mass
planets.  Microlensing's advantage for detecting large, distant planets
from the ground and compiling meaningful statistics therefrom
decreases as the time span for the radial velocity programs with the
highest precision increases.  However, most of the lenses will be M
stars, whereas these stars comprise only a small fraction of the
radial velocity targets.  In addition, the M stars will be located
primarily in the Galactic bulge in contrast to the local nature of the
stars in radial velocity searches. In that sense, microlensing
searches from the ground will always have a unique contribution to
make in constraining the statistical distribution of planets.  

Its unique sensitivity to
Earth mass planets may be less important because of the reduced
probability of detecting such planets from the ground.  This
importance would be drastically increased with the promise of an
enormous increase in the number of events and the continuous
monitoring of all the events with high time resolution, high precision
photometry from space (Bennett and Rhie, 2000).  Such a space based 
planet detection scheme using microlensing, if realized, would
decrease the motivation for ground based searches because of the drastically
increased throughput and higher sensitivity.  However, no such mission
has yet been funded, so it 
is premature to abandon continued planning for extensive ground based
microlensing searches for planets.  The robust nature of the large
probabilities of detection for Jupiter to perhaps Uranus mass planets
during an event demonstrated here supports the continuation of this
effort.  The need for the statistics of the occurrence of planets
distant from their sources that can be obtained relatively rapidly
from the ground only with a dedicated microlensing search will become
more urgent as planning for various space-based planetary searches
continues. 
\begin{center}{\bf Acknowledgements}\end{center}
\acknowledgements
It is a pleasure to thank Alex Filippenko, John Huchra, Penny Sackett
and especially Tim Sasseen for information and instruction concerning
photometric signal to noise ratios, Thanks also to Man Hoi Lee who
read the entire manuscript and offered many suggestions for 
improvements. Thanks are due an unknown referee whose careful review
and suggestions greatly improved the manuscript. Partial  support for
this work came from the NASA OSS and PG\&G Programs under grants
NAG5-7177 and NAG5-3646.  
\appendix
\section{The luminosity function \label{sc:lf}}
Zoccali {\it et al.} (2000) assemble a luminosity function (LF) for
stars in the Galactic bulge in $J$-band from their own data for stars
later than the main sequence turnoff and from Tiede {\it et al.} 
(1995) and Frogel \& Whitford (1987) for the giants. 
We approximate the LF in their Fig. 10 with three straight line segments and
further normalize the integral over the LF to unity such that 
\begin{eqnarray}
f_{\s L}(J)&\!\!\!=\!\!\!&2.628\times 10^{-8}e^{0.748\,J},\quad 9\le J\le
17,\nonumber\\
&\!\!\!=\!\!\!&1.108\times 10^{-14}e^{1.612\,J},\quad 17\le J\le 18,\nonumber\\
&\!\!\!=\!\!\!&6.096\times 10^{-5}e^{0.366\,J},\quad 18\le J\le 24 \label{eq:flj}
\end{eqnarray}        
represents the fraction of sources per unit $J$ magnitude.  Although
the extinction is lower in $J$, the sky
noise may be too excessive in this band for either the initial microlensing
survey or the high time resolution followup monitoring in search of
planetary perturbations (C. Stubbs, private communication 1999). We
will therefore convert the LF to $I$-band, 
for which we need $I-J$ as a function of $J$.  For this purpose, we
use Tables II and III of Bessell and Brett (1988), which give the
colors of dwarfs (Class V) and giants (Class III) respectively as a
function of spectral type in  terms of magnitude differences. 

$V-J$ and $I-J$ are determined as functions of $M_{\s V}$ with appropriate
relations among the columns in the tables, and $M_{\s V}$ is
substituted for the spectral types separately for dwarfs  and
giants. $M_{\s V}$ values are taken from $\S$ 96 of Allen
(1973). Linear fits to the points in the $V-J,\,M_{\s V}$ and
$I-J,\,M_{\s V}$ planes yield  
\begin{eqnarray}
V-J&=&0.361\,M_{\s V}-0.422\quad {\rm for\;dwarfs},\nonumber\\
&=&-1.833\,M_{\s V}+2.667\quad{\rm for\;giants},\label{eq:vjvsmv}\\
I-J&=&0.141\,M_{\s V}-0.159\quad{\rm for\;dwarfs},\nonumber\\
&=&-0.556\,M_{\s V}+0.944\quad{\rm for\;giants}.\label{eq:ijvsmv}
\end{eqnarray}
Since $V-J=M_{\s V}-M_{\s J}$ we have $M_{\s V}$ as a function of
$M_{\s J}$ which when substituted into the $I-J$ relations with
$M_{\s J}=J-14.5$, where 14.5 is the distance modulus for the center
of the Galaxy, yields $I-J$ as a function of $J$:
\begin{eqnarray}
I-J&=&0.221\,J-3.456 \quad {\rm for\;dwarfs},\nonumber\\
&=&-0.196\,J+3.263\quad {\rm for\;giants}. \label{eq:ijvsj}
\end{eqnarray}
We solve each of Eqs. (\ref{eq:ijvsj}) for
$J(I)$.  Then the LF in $I$-band is given by $f_{\s L}(I)=f_{\s
L}[J(I)](dJ/dI)$.  However, the range of $I$ for each segment differs
from the range in $J$;  $J=9,\,17,\,18,\,24\rightarrow I=
10.60,\,16.93,\,18.52,\,25.85$.  We have assumed the expressions for the
giants apply for $9\le J\le 17$ and the expressions for the dwarfs
apply for $18\le J\le 24$.  We round the values of $I$ corresponding
to the segment boundaries in $J$ slightly and assume the giant
relations apply for $10.5\le I\le 17$ and the dwarf relations for
$18.5\le I\le 26$.  The giant and dwarf segments are connected with a
straight line in the $I$-band equivalent of Fig. 12 of Zoccali {\it et
al.} (1999) and the LF renormalized so that
\begin{eqnarray}
f_{\s L}(I)&\!\!\!=\!\!\!&1.470\times 10^{-9}e^{0.930\,I_{\s 8}},\;\; 10.5\le
I_{\s 8}\le 17 \nonumber\\ 
&\!\!\!=\!\!\!&2.575\times 10^{-8}e^{0.762\,I_{\s 8}},\quad 17\le
I_{\s 8}\le 18.5
\nonumber\\ 
&\!\!\!=\!\!\!&1.316\times 10^{-4}e^{0.300\,I_{\s 8}},\; 18.5\le I_{\s 8}\le 26
\label{eq:lfi} 
\end{eqnarray}
represents the fraction of the sources per unit $I$ magnitude. (The
subscript stands for the 8 kpc distance to the Galactic center.)
The slope of the $I$-band luminosity function of Holtzman {\it et al.}
(1998) for the dim stars  is in reasonable agreement with the last of
Eqs. (\ref{eq:lfi}). 

The luminosity function will be used in two ways in the determination
of planetary detection probabilities:  First in the $S/N$
criterion for detection and second in the visibility of the sources.
Both will involve a modification of the luminosity
function to account for extinction as a function of the distance to
the source. The extinction $A_{\s I8}$ to 8 kpc
is about 0.76 magnitudes in Baade's window (Holtzman {\it et al.}
1998) but increases to about 18 toward the Galactic center.  $A_{\s
I8}$ will be retained as a free parameter 
to be adjusted for each LOS.  We assume the extinction is
uniformly distributed along the LOS such that the fractional
change in intensity $d{\cal I}/{\cal I}=-CdD_{\s OS}$ and
${\cal I}={\cal I}_0e^{-CD_{\s OS}}$, where 
$D_{\s OS}$ is the distance to the source and $C$ is constant.  Then
$e^{-CD_{\s OS}}=10^{-0.4A_{\s I}}=e^{-0.921A_I}$ yields
$CD_{\s OS}=0.921A_{\s I}=CD_8\zeta=0.921A_{\s I8}\zeta$  or the extinction
$A_{\s I}=A_{\s I8}\zeta$. The apparent magnitude of
a star at arbitrary $\zeta$ along a LOS where the extinction
to 8 kpc is $A_{\s I8}$ is then
\begin{equation}
I=I_8+A_{\s I8}\zeta+5\log{\zeta}. \label{eq:iappar}
\end{equation} 

For the visibility of the sources it is convenient to express the
luminosity function in terms of luminosity instead of magnitudes, so we
proceed to determine the $I$-band luminosities of the bulge stars.  
The $I$-band is centered near 0.80 $\mu$m, with a flux density at the
Earth from a 0 magnitude star like Vega of $1.20\times 10^{-12} \rm
W\,cm^{-2}\mu m^{-1}$ (McLean, 1997). The Gunn $I$ filter spans
0.75 to 0.88 $\mu m$ (FWHM) 
(http://ftp.noao.edu/kpno/filters/4inch\_list.html), so the flux density
in $I$ band  from Vega is $1.56\times 10^{-6}{\rm
ergs\,cm^{-2}sec^{-1}}$.  Then,
\begin{equation}
\frac{L_{\s I}}{L_{\s\odot}}=3.06\times 10^610^{-0.4I_{\s 8}}=
3.06\times 10^6e^{-0.921I_{\s 8}} \label{eq:li}
\end{equation}
is the luminosity of a star at the Galactic center whose $I$ magnitude is
$I_8$. 

The fraction of stars with $L_{\s I}>L_{\s I0}$ is the fraction of stars with
$I_{\s 8}<I_{\s 80}$, which is found by integrating Eq. (\ref{eq:lfi}) from
the minimum $I_8$ value of 10.5 to $I_{\s 80}$. The latter fraction is
\begin{eqnarray}
F(I_{\s 8}<I_{\s 80})&=&1.581\times 10^{-9}e^{0.930I_{\s 80}}-
2.75\times 10^{-5}\nonumber\\
&& 10.5\leq I_{\s 80}\leq 17\nonumber\\
&=&3.381\times 10^{-8}e^{0.762I_{\s 80}}-2.590\times 10^{-3}\nonumber\\
&&17\leq I_{\s 80}\leq 18.5\nonumber\\
&=&4.387\times10^{-4}e^{0.300I_{\s 80}}-0.0711\nonumber\\
&&18.5\leq I_{\s 80}\leq 26 \label{eq:fracili0}
\end{eqnarray}
Using Eq. (\ref{eq:li}) to convert $I_8$ to $L_{\s I}$, we find the
fraction of stars with $L_{\s I}>L_{\s I0}$ to be 
\begin{eqnarray}
F(L_{\s I}>L_{\s I0})&=&5.64\times 10^{-3}L_{\s
I0}^{-1.01}-2.75\times 10^{-5},\nonumber\\
&&194>L_{\s
I0}>0.487,\nonumber\\
&=&7.84\times 10^{-3}L_{\s I0}^{-0.827}-2.59\times 10^{-3},\nonumber\\
&&0.487>L_{\s I0}>0.123,\nonumber\\ 
&=&5.71\times 10^{-2}L_{\s I0}^{-0.326}-0.0711,\nonumber\\
&&0.123>L_{\s I0}>1.23\times 10^{-4}, \label{eq:flgtl0}
\end{eqnarray}
where $L_{\s I}$ and $L_{\s I0}$ are in units of $L_{\s\odot}$.
 
\noindent{\bf Source Visibility}

We modify the cumulative luminosity function of Eqs. (\ref{eq:flgtl0})
with extinction to determine the visibility of the sources to be
monitored in a microlensing survey. The flux density
received from a source at $\zeta$ is 
\begin{equation}
{\cal F }=\frac{L}{4\pi D_8^2\zeta^2}e^{-0.921A_{\s
I8}\zeta}.\label{eq:extinction} 
\end{equation}
The fraction of sources with $L_{\s I}>L_{\s I0}$ in Eqs. (\ref{eq:flgtl0})
has the form $C_1L_{\s I0}^\beta+C_2$.  
If ${\cal F}_{\s 0}$ is the minimum flux density that can be reliably
monitored in a planet search from  a source at $D_{\s OS}$, the cumulative
luminosity function takes the form
\begin{eqnarray}
F({\cal F}>{\cal F}_{\s 0})&=&\xi({\cal F}_0,\zeta,A_{\s
I8}) \label{eq:fracvis}\\
&=&C_{\s 1}\left(4\pi D_{\s 8}^2 {\cal F}_{\s 0}\zeta^2e^{0.921A_{\s
I8}\zeta}\right)^\beta +C_2. \nonumber
\end{eqnarray}  
Eq. (\ref{eq:fracvis}) represents the fraction of sources that are
visible at distance $\zeta$.

Although beyond current practice (P. Sackett, private communication,
2000), we shall assume a source with $I=21$ can be monitored
successfully from the ground in planet searches. This corresponds to
the minimum usable flux density of ${\cal F}_{\s 0}\rightarrow {\cal
F}_{\s 21}= 6.21\times 10^{-15}{\rm 
ergs\,cm^{-2}sec^{-1}}$. For the $I$ band LF that we have adopted, about
1.16\% of the stars have $I_{\s 8}<17$ and about 4.18\% have $I_{\s
8}<18.5$.  These are the points in the LF where the slope changes, so
the values of $C_1,\,C_2,\,\beta$ must change for $F$ in
Eq. (\ref{eq:fracvis}) when $F=0.0116$ and 0.0418.  Of course
$F=1$ for sufficiently small values of $\zeta$ where all of the stars
are visible, and $F=0$ for sufficiently large values of $\zeta$ where
no stars are visible, where ``visible'' here means $I<21$.  For
example, substitution of 
the values of the constants from Eqs. (\ref{eq:flgtl0}) along with
the value of ${\cal F}_{\s 21}$ into Eq. (\ref{eq:fracvis}) yields
\begin{eqnarray}
F({\cal F}>{\cal F}_{\s 21})&=&\xi({\cal F}_0,\zeta,A_{\s
I8})\nonumber\\
&=&1,\qquad\qquad\zeta<0.0892,\nonumber\\
&=&0.240\left(\zeta^2e^{0.921A_{\s
I8}\zeta}\right)^{-0.326}-0.0711,\nonumber\\
&&0.0892<\zeta<0.9075,\nonumber\\
&=&0.300\left(\zeta^2e^{0.921A_{\s
I8}\zeta}\right)^{-0.827}-.000259,\nonumber\\ 
&&0.9075<\zeta<1.2025, \nonumber\\
&=&0..482\left(\zeta^2e^{0.921A_{\s I8}\zeta}\right)^{-1.010}
-.0000275,\nonumber\\ 
&&1.2025<\zeta<2.7677,\nonumber\\
&=&0,\qquad\qquad 2.7677<\zeta \label{eq:Fforai=3}
\end{eqnarray}
as the fraction of sources visible at distance $\zeta$ for $A_{\s
I8}=3$. For $A_{\s I8}=1$ the expressions are identical to those in
Eqs. (\ref{eq:Fforai=3}) except the breaks occur at $\zeta=0.0965,\,
1.5536,\,2.2477,\,$ and 6.4602.  In this latter case the most luminous
star in the LF could be seen far beyond the opposite edge of the
Galaxy even if the absorption were to persist all along the line of
sight.  For $A_{\s I8}=5$ the breaks are at $\zeta=0.0833,\;0.6739,\; 
0.8648$ and 1.8384. Fig. \ref{fig:visibilityi21} shows the fraction of visible
sources ($I<21$) as a function of $\zeta$ for three values of $A_{\s I8}$.
\section{Amplification of the source \label{sc:appenA}}
We adopt the analysis of Witt (1990) to determine the amplification of
a point source star (source) by the star-planet binary lens as a
function of the 
relative angular positions of the source, lensing star of mass $M$
(hereinafter, the lens) and the planet $m$.  However, since $M\gg
m$, we choose the optical axis to pass through the lens and locate
source and planet relative to this axis. All displacements of the
projected positions on the lens plane are normalized by the Einstein
ring radius of $M$, $R_{\s E}=\sqrt{4GMD_{\s OL}(D_{\s OS}-D_{\s
OL})/(c^2D_{\s OS})}$, where $G$ is the gravitational constant, $c$ is
the velocity of light and $D_{\s OL}$ and $D_{\s OS}$ are the
observer-lens and observer-source distances respectively.  ($R_{\s E}$
is the radius of the axially symmetric ring image of a source that is
directly behind the lens.) The lens
equation is then written
\begin{equation}
\lambda=w-\frac{1}{w^*}+\frac{\epsilon}{\rho^*-w^*},\label{eq:lenseq}
\end{equation}
where $\lambda=\xi+i\,\eta$ and $w=x+i\,y$ are the displacements of the
source and image from the optical axis respectively,
$\rho=\rho_x+i\,\rho_y$ is the displacement of the planet and
$\epsilon=q$. The (*) indicates complex conjugation.
$\lambda,w,\rho$ may be interpreted as either the projected
displacements in the lens plane normalized by $R_{\s E}$ or the 
angular displacements normalized by the angle subtended by $R_{\s E}$.
The axes ($\xi,\eta$) and ($x,y$) are aligned.

If we multiply the complex conjugate of Eq. (\ref{eq:lenseq}) by the
product $w(\rho-w)$ and substitute $w$ as a function of $w*$ from
Eq. (\ref{eq:lenseq}), the following polynomial results:
\begin{equation}
P(w^*)=a_5{w^*}^5+a_4{w^*}^4+a_3{w^*}^3+a_2{w^*}^2+a_1{w^*}+a_0=0,
\label{eq:wstarpolynom}
\end{equation}
where
\begin{eqnarray}
a_5&=&\lambda(\lambda-\rho) \nonumber \\
a_4&=&\lambda(\lambda(-\lambda^*-2\rho^*)+\rho\lambda^*+2\rho\rho^*
+1+\epsilon)-\epsilon\rho \nonumber \\
a_3&=&\lambda^*(\lambda(2\rho^*\lambda-2(1+\epsilon)-2\rho\rho^*)+
\rho(1+\epsilon))\nonumber\\
&&+\lambda(\lambda{\rho^*}^2-\rho{\rho^*}^2-2\rho^*) +
\epsilon\rho\rho^* \nonumber \\
a_2&=&\lambda^*(\lambda(-{\rho^*}^2\lambda+\rho^*(\rho\rho^*+2\epsilon+4))
-(2+\epsilon)\rho\rho^*\nonumber \\
&&-(1+\epsilon)^2)+\lambda{\rho^*}^2(1-\epsilon)+
\rho^*\epsilon(1+\epsilon) \nonumber \\
a_1&=&\lambda^*(-2{\rho^*}^2\lambda+\rho^*(\rho\rho^*+2(1+\epsilon)))-
\epsilon{\rho^*}^2 \nonumber \\
a_0&=&-\lambda^*{\rho^*}^2. \label{eq:ccoeff}
\end{eqnarray}

The roots of Eq. (\ref{eq:wstarpolynom}) are the complex conjugates of
the positions of the images, where the number of valid images is
either 3 or 5 with validity determined by substitution of the
corresponding root into Eq. (\ref{eq:lenseq}).

Eq. (\ref{eq:lenseq}) represents a transformation of the coordinates
of an image ($x,y$) to the coordinates of the source ($\xi,\eta$).
The amplification of the source represented by a particular image is
the ratio of an element of area at the image position in the ($x,y$)
system, as transformed from the source position in the ($\xi,\eta$)
system, to the original element in the ($\xi,\eta$) system. Hence, the
total amplification of the source due to the star-planet binary lens
is
\begin{equation}
A=\sum_i|{\rm det}\,J_i|^{-1},\label{eq:amp}
\end{equation}
where (Witt, 1990)
\begin{eqnarray}
{\rm det}(J)&=&\frac{\partial\xi}{\partial x}\frac{\partial\eta}
{\partial y}-\frac{\partial\xi}{\partial y}\frac{\partial\eta}{\partial
x}\nonumber\\
&=& \left(\frac{\partial\lambda}{\partial w}\right)^2-\frac{
\partial\lambda}{\partial w^*}\left(\frac{\partial\lambda}{\partial
w^*}\right)^*\nonumber\\
&=&1-{\vrule\frac{1}{{w^*}^2}+\frac{\epsilon}{(\rho^*
-w^*)^2}\vrule\,}^2  \label{eq:detj}
\end{eqnarray}
is the determinant of the Jacobian of the transformation represented
by Eq. (\ref{eq:lenseq}), and $J_i=J(w_i)$ is the Jacobian evaluated at the
coordinates of the $i$th image.  The sum is over either three or five
images.  For a single lens without the planet the amplification
simplifies to 
\begin{equation}
A_0=\frac{u^2+2}{u\sqrt{u^2+4}}, \label{eq:a0}
\end{equation}
where $u$ is the angular separation of the lens and source normalized
by the angle subtended by the Einstein ring radius in the lens plane.

The curves defined in $w$ space by $\det{J}=0$ are called the critical
curves where the amplification is formally infinite.  This singularity
is removed when the finite size of the source is considered. The
corresponding closed curves in $\lambda$ space found from
Eq. (\ref{eq:lenseq}) are the caustics, such that when the source is 
on a caustic curve an image is on a critical curve.  For the single
star lens, the caustic reduces to a single point at the lens position,
and the corresponding critical curve is the Einstein ring. If the
source never crosses a caustic during the event there are always just
three images and therefore three terms on the right-hand-side of
Eq. (\ref{eq:amp}). At the first caustic crossing, an image is created
on the critical curve and separates into two images for a total of
5 images that persist as long as the source is inside the closed
caustic curve.  On the second caustic crossing, with the source now
outside the closed caustic, one of images created at the first
crossing merges with one of the original three images and
disappears---leaving again only three images.

To evaluate contours of equal amplification for a given position of
the source, the amplification is evaluated for a $401\times 401$ grid
of planet positions centered on the lens and extending $\pm 2.5 R_{\s
E}$ in both orthogonal directions.  A $401\times 401$ array of the
fractional change in the amplification due to the planet, $A/A_0-1$,
is saved, where $A$ is the amplification for a 
particular position of the planet and $A_0$ is the unperturbed
amplification for the given source position. Contours of constant
fractional change in the amplification (the perturbation due to the
planet) are obtained with any contouring program for arbitrary
fractional changes.  In Fig. \ref{fig:contoursweepiband}, the contours
shown correspond to the assumed minimal perturbation that can be
detected according to the $S/N$ at that position on the
light curve.  However, these contours are not used to trace their
extremes along the line containing the source and lens during the
course of the event.

It is more efficient to determine these extremes directly by moving
the planet along the axis containing the source and lens to determine
the two points on opposite sides of each unperturbed image position
where the fractional change in the amplification is $2/(S/N)$ for the
major image outside $R_{\s E}$ and $-2/(S/N)$ for the minor image
inside $R_{\s E}$.  The lens-source line
is chosen to be the real axis so that $\lambda=\lambda^*$ and
$\rho=\rho^*$. The coefficients in Eq. (\ref{eq:ccoeff})  are then
real and analytic in $\rho$, which is now the 
linear position of the planet along the real axis.  We form the
function 
\begin{equation}
f(\rho)=\frac{A}{A_0}-\frac{1}{A_0}\sum_i |{\rm det}(J_i)|^{-1},
 \label{eq:frho}
\end{equation}
where $A=A_0[1\pm 2/(S/N)]$ defines the extremes of the planet positions
along the real axis on either side of the unperturbed image positions
inside of which the fractional perturbation would exceed $2/(S/N)$ and allow
detection of the planet.  These extremes are defined by the zeros of
$f(\rho)$.  A trial value of $\rho$ somewhat removed from the
unperturbed position of a real image is used to start a ``safe''
Newton's method solution for a zero of $f(\rho)$ (Press, {\it et al.}
1986).  All the necessary
derivatives can be found from Eqs. (\ref{eq:wstarpolynom}) and
(\ref{eq:ccoeff}), where $\rho^*=\rho$ in the latter, and the realness
of the coefficients is exploited.  A step in the procedure is thus for
a trial position of $\rho$, all of the valid roots of the polynomial form of
the lens equation (Eq. (\ref{eq:wstarpolynom})) are used to form
$f(\rho)$ and $df(\rho)/d\rho$ for use in Newton's method to determine
the next approximation to the root, where the procedure is iterated
until a tolerable precision is obtained.  The two pairs of roots are
determined for each position of the source as it is stepped through
the event, where $S/N$ is a function of the source position.
\section{Probability distribution of projected star-planet
separation \label{sc:appenB}}
If $dn_{\s L}(M,D_{\s OL})/dM$ is the number density of stars (lenses)
per unit mass interval near mass $M$ and near distance $D_{\s OL}$,
then $(dn_{\s L}(M,D_{\s OL})/dM)D_{\s OL}^2dD_{\s OL}dM\Delta\Omega$
is the number of stars in mass range $dM$ about $M$ and in volume
$D_{\s OL}^2dD_{\s OL}\Delta\Omega$, 
where $\Delta\Omega$ is a representative solid angle in the telescope
field of view.  The fraction of the lenses in the volume of those
along the LOS to the source is just this quantity divided by
its integral along the LOS and over the mass
function, where we assume the MF to be spatially invariant. Hence, 
\begin{displaymath}
F(x_p,a,\zeta)=
\end{displaymath}
\begin{displaymath}
\displaystyle\int_0^{\zeta}\int_{M_{min}}^{M_{max}}F(x_p,a,\zeta,z,M)
\Theta\left[\frac{a^2\zeta}{KMz(\zeta-z)}-x_p^2\right]\times
\end{displaymath}
\begin{displaymath}
z^2\frac{dn_{\s L}(M,z)}{dM}dMdz\bigg/
\end{displaymath}
\begin{equation}
\displaystyle \int_0^\zeta
\int_{M_{min}}^{M_{max}}z^2\frac{dn_{\s L}(M,z)}{dM}dMdz
\label{eq:fxpave}
\end{equation} 
is the average of $F(x_p,a,\zeta,z,M)$ over the lens distribution from
observer to source and over the MF.  We have normalized the distances
with $D_8$ as above and introduced the step function $\Theta(x)$ to
exclude those regions of $z,M$ space for which the integrand in the
numerator is undefined. If $x_p^2>a^2\zeta/(KMz(\zeta-z))$, $r>a$ in
Eq. (\ref{eq:cumprobfxp}).  For fixed $x_p$ and $a$,
Eq. (\ref{eq:fxpave}) represents the average of $F(x_p,a,\zeta,z,M)$
over those lenses distributed along the LOS to distance $\zeta$ for
which $R_{\s E}<a/x_p$. 

For the density distribution in the Galaxy  we adopt the double
exponential model for the disk distribution of mass and a triaxial bar
distribution for the bulge. The disk model is that of Bahcall and
Soneira (1980), 
\begin{equation}
\rho_{\s M}=\int_{M_{\s min}}^{M_{\s max}}
\frac{d\rho}{dM}dM=\rho_{\s
M0}\exp{\left(\frac{-|z^{\s\prime}|}{300{\rm pc}}-\frac{r}{s_{\s
d}}\right)},   \label{eq:disk}
\end{equation}
where $z^{\s\prime}$ is the coordinate perpendicular to the plane of
the galaxy, $r$ is the radial coordinate in the plane of the galaxy, and
$s_{\s d}=2.7$ kpc is the scale length in the radial direction, which is
chosen less than the 3.5 kpc used by Bahcall and Soneira (Zhao, Spergel \&
Rich 1995; Kent, Dame \& Fasio 1995), and where $\rho_{\s M0}$ is
chosen such that $\rho_{\s M}=0.05M_{\s\odot}/{\rm pc^3}$ at
$z^{\s\prime}=0$ and $r=8$ kpc.  Like Zhao, Rich and Spergel (1996), we
choose the triaxial bulge model of Zhao (1996) but keep the nucleus in
a truncated form and do not terminate the bulge at 3.3 kpc:
\begin{equation}
\rho_{\s M}=\rho_{\s 0}\left[\exp\left(-\frac{s_{\s b}^2}{2}\right)+s_{\s
a}^{-1.85}\exp(-s_{\s a})\right], \label{eq:bulge}
\end{equation}
where 
\begin{eqnarray}
s_{\s b}^4&=&\left[\left(\frac{x}{\sigma_{\s x}}\right)^2+\left(\frac{y}
{\sigma_{\s y}}\right)^2\right]^2+\left(\frac{z^{\s\prime}}{\sigma_{\s
z}}\right)^4,\cr
s_{\s a}^2&=&\frac{q_{\s a}^2(x^2+y^2)+z^{{\s\prime}2}}{\sigma_{\s
z}^2}\quad {\rm if}\;\sqrt{x^2+y^2}>0.56\,{\rm kpc},\cr
&=&\frac{0.3136q_a^2+z^{{\s\prime}2}}{\sigma_z^2}\quad {\rm if}\;
\sqrt{x^2+y^2}\leq 0.56\,{\rm kpc} \label{eq:blgexp} 
\end{eqnarray}
with $q_{\s a}=0.6$, $\sigma_{\s x}=1.49$ kpc, $\sigma_{\s y}=0.58$ kpc,
and $\sigma_{\s z}=0.40$ kpc. The coefficient $\rho_{\s
0}$ determines the mass of the bulge.  The nucleus is truncated to
avoid the singularity at the origin in the Zhao bulge model. The
observations are consistent with the long axis of the bulge inclined
about $13^\circ$ to $20^\circ$ relative to the LOS to the
Galactic center with the near side of the bar lying in the second
quadrant.  We shall choose $\theta=-13^\circ$, but note that this
angle is very uncertain.  

The origin of coordinates for both disk and bulge distributions is
transferred from the Galactic center to the position of the
observer with the following transformation:
\begin{eqnarray}
x&=&\cos{\theta}-z\cos{b}\cos{(\ell-\theta)},\nonumber\\
y&=&-\sin{\theta}-z\cos{b}\sin{(\ell-\theta)},\nonumber\\
z^{\s\prime}&=&z\sin{b},  \label{eq:blgtrans}
\end{eqnarray}
for a lens in the bulge where $z$ is the distance along the line of
sight normalized by $D_{\s 8}$ with $xyz^{\s\prime}$ being galactocentric
coordinates  along the bulge principal axes normalized by $D_8$ and
$\theta$ the inclination of the bulge axis to the line between the
observer and the Galactic center  measured in the direction of 
increasing $\ell$.  For the disk distribution, the transformation is
\begin{eqnarray}
r^2&=&1+z^2\cos^2{b}-2z\cos{b}\cos{\ell},\nonumber\\
z^{\s\prime}&=&z\sin{b}, \label{eq:disktrans}       
\end{eqnarray}

For a dynamically constrained bulge mass of $2.2\times
10^{10}M_{\s\odot}$ (Zhao, 1996), $\rho_{\s 0}$ in Eq. (\ref{eq:bulge})
is $2.5\,M_{\s\odot}{\rm pc^{-3}}$.  The coefficient
$\rho_{\s M0}=0.05e^{8/s_d}M_{\s\odot}\,{\rm pc^{-3}}$ ($s_d$ in kpc) in
Eq. (\ref{eq:disk}) to satisfy the constraint on the local mass
density.  The conversion of these mass distributions to number density
distributions for use in $dn_{\s L}/dM$ depends on the assumed mass
function.   We shall adopt the Holtzman {\it et al.} (1998) MF
for both disk and bulge stars given by 
\begin{eqnarray}
\phi(M)&=&0.7795M^{-1}\qquad 0.08\leq M\leq 0.7 \nonumber\\
&=&0.5081M^{-2.2}\qquad 0.7\leq M\leq 2.0. \label{eq:holtzman}
\end{eqnarray}
The MF in Eq. (\ref{eq:holtzman}) is defined so that
$M\phi(M)dM$ represents the mass of stars/{$\rm pc^3$} in mass range
$dM$ about $M$ normalized such that the total mass density is
$1M_{\s\odot}{\rm pc^{-3}}$. A more recent MF  (Zocalli {\it et al.}
2000) has an index of 1.3 instead of 1 in the M star region.  However,
we use the Holtzman {\it et al.} MF and point out the effects
of increasing the index for small mass stars in
Fig. \ref{fig:pavemfcmpre}.  We truncate the MF
at $2M_{\s\odot}$ because of the small number of stars of
higher mass.  We could extend the MF into the brown dwarf
region, but the microlensing event time scale frequency distribution
does not seem compatible with a large number of brown dwarfs---at least
in the Galactic bulge (Peale, 1999). 

The number of stars/$M_{\s\odot}$, or equivalently, stars/${\rm pc^3}$ for
this assumed MF is found from integration of Eq. (\ref{eq:holtzman})
to be  2.1562 stars/$M_{\s\odot}$, so the number density of stars
corresponding to the mass densities given by Eqs. (\ref{eq:disk})  and
(\ref{eq:bulge}) are found by multiplying each by this factor.
The expression for $dn_{\s L}/dM$ is the sum of the disk and bulge
contributions at a given point, so
\begin{eqnarray}
&dn_{\s L}(M,z)/dM=\nonumber\\
&0.7795[0.1078e^A+5.3903(e^{-\frac{s_b^2}{2}}+s_a^{-1.85}
e^{-s_a})]/M,\nonumber\\
& 0.08<M<0.7\nonumber\\
&=0.5081[0.1078e^A+5.3903(e^{-\frac{s_b^2}{2}}+s_a^{-1.85}
e^{-s_a})]/M^{2.2},\nonumber\\
& 0.7<M<2.0 \label{eq:dnlz}
\end{eqnarray}
where $A$ is the exponent given  in Eq. (\ref{eq:disk}) $+8/s_d$ but
with the change of coordinates given in Eqs. (\ref{eq:disktrans}) and
(\ref{eq:blgtrans}) in all of $A,\,s_a,\,{\rm and}\,s_b$.  As we shall be 
dealing in fractional distributions, only the ratio of the
coefficients of the disk and bulge distributions will have any effect
on the probabilities. The integrals in Eq. (\ref{eq:fxpave}) are
evaluated with a Monte Carlo technique (Press {\it et al.} 1986) as a
function of $x_p$ for given $a$. 
\begin{center}{\bf References}\end{center}
\parindent=0pt
\parskip=10pt
Albrow M, Beaulieu JP, Birch P, Caldwell AR, Kane S. {\it et
al.} (1998) The 1995 pilot campaign of PLANET: Searching for
microlensing anomalies through precise, rapid round-the-clock
monitoring. {\it Astrophys. J.} {\bf 509}, 678-702. 

Albrow MD. Beaulieu J-P, Caldwell JAR, DePoy DL, Dominik M, Gaudi BS,
Gould A, Greenhill J, Hill K, Kane S, Martin R, Menzies J, Naber
RM, Pogge RW, Pollard KR, Sackett PD, Sahu KC, Vermaak P, Watson
R, Williams A. (2000a) Limits on Stellar and Planetary Companions in
Microlensing Event OGLE-1998-BUL-14, {\it Astrophys. J.} {\bf 535},
176-89.  

Albrow MD, An J, Beaulieu J-P, Caldwell JAR, Depoy DL, Dopminik M,
Gaudi BS, Gould A, Greenhill J, Hill K, Kane S, Martin R, Menzies J,
Naber RM, Pel J-W, Pogge RW, Pollard KR, Sackett PD, Sahu KC, VErmaak
P, Vreeswijk PM, Watson R, Williams A. (2000b) Limits on the abundance
of galactic planets from five years of PLANET observations,
astro-ph/0008078. {\it Astrophys. J.} In press.

Alcock C, Allsman RA, Alves DR, Axelrof TS, Becker AC, Bennett DP,
Cook KH, Drake AJ, Freeman KC, Geha AM, Griest K, Lehner MJ, Marshall,
SL, Minniti D, Nelson CA, Peterson BA, Popowski P, Pratt MR, Quinn PJ,
Stubbs CW, Sutherland W, Tromaney AB, Vandehei T, Welch DL. (2000) The
Macho project: Microlensing optical depth towards the galactic bulge
from difference image analysis, astro-ph 0002510. {\it Astrophys. J.}
In press.

Allen CW. (1973) {\it Astrophysical Quantities} Athone Press, Univ. of
London, p. 200. 

Bahcall JN, Soneira RM. (1980) The universe at faint
magnitudes. I. Models for the galaxy and the predicted star
counts. {\it Astrophys. J. Sup.} {\bf 44} 73-110.

Basu S. Rana NC. (1992) Multiplicity corrected mass function of
main-sequence stars in the solar neighborhood, {\it Astrophys. J.}
{\bf 393}, 373-84.

Beichman CA. (1996) (Ed.) A road map for the exploration of
neighboring planetary systems, publication 92-22, Jet Propulsion Lab.,
Pasadena, CA.

Bennett DP, Rhie SH (1996) Detecting Earth-Mass Planets with
Gravitational Microlensing. {\it Astrophys. J.} {\bf 472}, 660-64.

Bennett DP, Rhie SH (2000) The Galactic Exoplanet Survey Telescope: A
proposed space-based microlensing survey for terrestrial extra-solar
planets, Astro-Ph 0003102, In {\it Proceedings of the Disks,
Planetesimals \& Planets Meeting, Tenerife, Jan. 24-28, 2000} In press.

Bessell MS, Brett JM. (1988) JHKLM photometry: Standard systems,
passbands, and intrinsic colors, {\it Pub. Astron. Soc. Pac.} {\bf
100}, 1134-51.

Charbonneau D, Brown TM, Latham DW, Mayor M. (2000) Detection of
planetary transits across a Sun-like star. {\it Astrophys. J.} {\bf
529}, L45-L48.

Frogel JA, Whitford AE. (1987) M giants in Baade's window - Infrared
colors, luminosities, and implications for the stellar content of E
and S0 galaxies, {\it Astrophys. J.} {\bf 320}, 199-237.  

Gatewood G. (1991) The Allegheny Observatory search for planetary
systems, {\it IAF, International Astronautical Congress, 42nd},
Montreal, Canada Oct. 5-11.

Gaudi BS, Sackett PD. (2000) Detection efficiencies of microlensing
datasets for stellar and planetary companions, {\it Astrophys. J.} {\bf
528}, 56-73.

Gould A, Loeb A (1992) Discovering planetary systems through
gravitational microlensing, {\it Astrophys. J.} {\bf 396}, 104-14.

Gould A, Bahcall J, Flynn C. (1997) M dwarfs from Hubble Space
Telescope star counts. III. The Groth Strip, {\it Astrophys. J.} {\bf
482}, 913-18.

Griest K, Safizadeh N. (1998) The use of high-magnification
microlensing events in discovering extrasolar planets. {\it
Astrophys. J.} 500, 37-50.

Henry GW, Marcy GW, Butler RP, Vogt SS. (2000) A transiting ``51
Peg-like'' planet, {\it Astrophys J.} {\bf 529}, L41-L44

Holtzman JA, Watson AM, Baum WA, Grillmair CJ, Groth EJ, Light RM,
Lynds R, O'Neil EJ. (1998) The luminosity function and initial mass
function in the Galactic bulge, {\it Astron. J.} {\bf 115}, 1946-57.

Kent SM, Dame TM, Fasio G. (1991) Galactic structure from the spacelab
infrared telescope 2. Luminosity models of the milky way, {\it
Astrophys. J.} {\bf 378}, 131-38.

Lattanzi MG, Spagna A, Sozzetti A, Casertano S. (2000) {\it
Mon. Not. Roy. Ast. Soc.} In press.

Mao S, Paczy\'nski B. (1991) Gravitational microlensing by double
stars and planetary systems, {\it Astrophys. J.} {\bf 374}, L37-L40. 

Marcy GW, Cochran WD, Mayor M. (2000) Extrasolar planets around
main-sequence stars, In {\it Protostars and Planets
IV} eds. V Mannings, AP Boss and SS Russel, Univ of Arizona Press,
Tucson AZ. p. 1285.

McLean I. (1997) {\it Electronic Imaging in Astronomy}, John Wiley and
Sons, New York, p304.

Peale SJ. (1997) Expectations from a microlensing search for planets,
{\it Icarus} {\bf 127}, 269-89.

Peale SJ. (1998) On microlensing event rates and optical depth toward
the galactic center, {\it Astrophys. J.} {\bf 509}, 177-91.

Peale SJ. (1999) Newly discovered brown dwarfs not seen in
microlensing timescale frequency distribution? {\it Astrophys. J.}
{\bf 524}, L67-L70.

Press WH, Flannery BP, Teukolsky SA, Vetterling WT. {\it Numerical
Recipes} Cambridge University Press, London, p. 221.

Rhie SH, Becker A, Bennett DP, Calitz J, Cook K, Fragile P, Johnson B,
Gurovich S, Han C, Hoffman M, Laws C, Martinez P, Meintjes P,  Minniti
D, Park S, Peterson B, Quinn J. (2000)  Microlensing Planet Search
Project, {\it Planetary systems in the universe, International
Astronomical Union. Symp no. 202} Manchester, England , August 2000.

Swain MR, Akeson RL, Colavita MM, Shao M. (2000) Detecting extrasolar
planets with the Keck interferometer, {\it Planetary systems in the
universe, International Astronomical Union. Symp no. 202} Manchester,
England , August 2000. 

Tiede GP, Frogel JA, Terndrup DM. (1995) Implications of new JHK
photometry and a deep infrared luminosity function for the Galactic
bulge, {\it Astron. J.} {\bf 110}, 2788.

Walker AR. (1987) {\it NOAO Newsletter \#10}, June.

Witt HJ. (1990) Investigation of high amplification events in light
curves of gravitationally lensed quasars. {\it Astron. Astrophys.}
{\bf 236}, 311-322.

Zhao HS, Spergel DN, Rich RM. (1995) Microensing by the galactic bar,
{\it Astrophys. J.} {\bf 440}, L13-L16.

Zhao HS, Rich RM, Spergel DN. (1996) A consistent microlensing model
for the galactic bar, {\it Mon. Not. Roy. Ast. Soc.} {\bf 282},
175-81. 

Zhao HS. (1996) A steady state dynamical model for the COBE-detected
galactic bar, {\it Mon. Not. Roy. Ast. Soc.} {\bf 283}, 149-66.

Zoccali M, Cassisi S, Frogel JA, Gould A, Ortolani S, Renzini A, Rich
RM, Stephens AW. (2000) The initial mass function of the galactic bulge
down to $\sim0.15 M_{\s\odot}$, {\it Astrophys. J.} {\bf 530}, 418-28.
\end{document}